%% file: article.tex
\documentclass[floatfix,prb,superscriptaddress,twocolumn,letterpaper,nobalancelastpage]{revtex4-2}

\usepackage{mycommands}

\begin{document}
    
    \title{Linear and Non-Linear Response of Quadratic Lindbladians}
        
    \author{Spenser Talkington}
    \email{spenser@upenn.edu}
    \affiliation{Department of Physics and Astronomy, University of Pennsylvania, Philadelphia, Pennsylvania 19104, USA}
    
    \author{Martin Claassen}
    \affiliation{Department of Physics and Astronomy, University of Pennsylvania, Philadelphia, Pennsylvania 19104, USA}
    
    \date{\today}
    \begin{abstract}
        Quadratic Lindbladians encompass a rich class of dissipative electronic and bosonic quantum systems, which have been predicted to host new and exotic physics. In this study, we develop a Lindblad-Keldysh spectroscopic response formalism for open quantum systems that elucidates their steady-state response properties and dissipative phase transitions via finite-frequency linear and non-linear probes. As illustrative examples, we utilize this formalism to calculate the (1) density and dynamic spin susceptibilities of a boundary driven XY model at and near criticality, (2) linear and non-linear optical responses in Bernal bilayer graphene coupled to dissipative leads, and (3) steady state susceptibilities in a bosonic optical lattice. We find that the XY model spin density wavelength diverges with critical exponent 1/2, and there are gapless dispersive modes in the dynamic spin response that originate from the underlying spin density wave order; additionally the dispersing modes of the weak and ultra-strong dissipation limits exhibit a striking correspondence since the boundary dissipators couple only weakly to the bulk in both cases. In the optical response of the Bernal bilayer, we find that the diamagnetic response can decrease with increasing occupation, as opposed to in closed systems where the response increases monotonically with occupation; we study the effect of second harmonic generation and shift current and find that these responses, forbidden in centrosymmetric closed systems, can manifest in these open systems as a result of dissipation. We compare this formalism to its equilibrium counterpart and draw analogies between these non-interacting open systems and strongly interacting closed systems.
    \end{abstract}

    \maketitle

    
    \section{Introduction}\label{section:intro}
        
        Quantum systems rarely exist in perfect isolation from their environment. While commonly regarded as detrimental for quantum information processing \cite{leon21}, the dynamics of open quantum systems have recently attracted substantial interest as a route towards harnessing engineered dissipation as a resource for creating exotic quantum states and information processing \cite{verstraete09,harrington22}. Tailored environmental couplings in materials and quantum devices have long been appreciated as a possible ingredient for entanglement generation and quantum computation \cite{hebenstreit19,shtanko21}. At the same time, the steady states of open systems can offer rich new prospects for stabilizing unconventional topological \cite{bardyn2013topology,yang21} and many-body states \cite{diehl08,buca18,tindall19} not readily found in thermal equilibrium, while posing new challenges in their theoretical and computational modeling \cite{weimer2021simulation}. To understand signatures of such dissipative states of matter in experiment, a key question hence concerns how to model and interpret dynamical and spectroscopic responses in open quantum systems.

        While open quantum systems are generically non-Markovian \cite{breuer2002theory,breuer2016colloquium,de2017dynamics} and can be simulated effectively in only a limited number of cases \cite{weimer2021simulation}, Markovian master equations and in particular the Lindblad master equation are prominent for their interpretibility, effectiveness as an approximation for steady-state behavior, and tractability for analytic work \cite{lindblad1976generators,gorini1976completely,plenio1998quantum,barthel2022solving}. Fundamentally, the Lindblad equation describes the time-local evolution of the density matrix of a small electronic or bosonic subsystem coupled to a larger Markovian reservoir, and has a natural field-theoretic representation via time-local self energies on the Keldysh contour \cite{sieberer2016keldysh,kamenev2023field,thompson2023field,mcdonald2023third,kim2023third}. For fermions and bosons, the Lindblad master equation moreover can be usefully represented in terms of ``third-quantized" superoperators which obey standard (anti)commutation relations \cite{prosen2008third,prosen2010exact,seligman2010third}. Remarkably, when such a Lindblad master equation is assembled from a non-interacting (quadratic) Hamiltonian and dissipative particle gain and loss processes, its eigenmodes can be computed exactly from diagonalizing a single particle matrix, which can then be used to construct the full many-body Lindbladian \cite{prosen2008third,prosen2010exact,seligman2010third}. This quadratic structure means that the Keldysh path integrals are Gaussian and hence Wick's theorem can be used to reduce multi-point functions to two point functions as will be key in this work.

        While Lindbladians originally originated from the desire to study individual quantum-optical emitters \cite{plenio1998quantum}, recently extended dissipative many-particle systems have attracted substantial attention.
        Particular model systems include boundary driven spin systems and quantum circuits that exhibit non-equilibrium phase transitions \cite{prosen2008quantum,eisert2010noise,carollo2020geometry,bertini2021finite,landi2022nonequilibrium}, and qubit systems where Lindbladians have been used to model dissipation \cite{albert2014symmetries} and realize dissipative versions of the Ising model on quantum computers \cite{mi2023stable}.
        Lindbladians have also been applied to reveal unusual dynamics in SYK systems \cite{kulkarni2022lindbladian,kawabata2023dynamical}, to realize dark states and dissipation-induced flat bands \cite{talkington2022dissipation}, study localization \cite{beck2021disorder,thompson2024localization}, and to stabilize entangled many-body states \cite{pocklington2022stabilizing,pocklington2023stability,pocklington2023dissipative}.
        In quantum circuits, Lindbladians have been used to complete logical operations \cite{shtanko2021complexity}, and even as part of a feedback mechanism to mitigate errors in quantum computing experiments \cite{van2023probabilistic}.
        Previous work on Lindbladian systems has focused on the structures of subspaces and symmetries \cite{buvca2012note,albert2014symmetries,talkington2022dissipation,muller2022measurement,mcdonald2022exact} as well as the response of these subsystems \cite{albert2016geometry}. Additionally, recent efforts have succeeded in classifying topological states in these systems \cite{lieu2020tenfold} and in non-Hermitian systems more broadly \cite{bergholtz2021exceptional}.
        Building on these rapid advances of new and emergent behaviors in dissipative quantum systems, a natural question to address is how to elucidate and interpret their properties using experimentally-accessible spectroscopic tools.

        In this paper, we develop a theory of linear and non-linear spectroscopy of the steady-state properties of Lindbladian quantum systems. For quadratic Lindbladians, we show that linear and non-linear spectroscopic responses have a succinct representation in terms of the biorthogonal eigenmodes of the third-quantized quadratic single-particle Lindbladian, and illustrate consequences of optical conductivities, second harmonic generation and shift currents of electronic Lindbladians, as well as dynamical spin correlation functions of boundary-driven spin chains.
        The paper is organized as follows: In Section \ref{sec:GFs}, we start by reviewing Lindblad-Keldysh Green's functions and third quantization, and compute the single-particle Green's functions for quadratic Lindbladians. In Section \ref{sec:response} we derive expressions for the linear and non-linear response of fermionic and bosonic open quantum systems governed by a quadratic Lindbladian; finally, in Section \ref{sec:examples} we apply our formalism to calculate response properties for (1) a 1D XY spin chain, (2) Bernal bilayer graphene, and (3) a bosonic optical lattice.

    \section{Results}

    \subsection{Lindblad Keldysh Green's Function Formalism}\label{sec:GFs}

        We consider the time evolution of the density matrix $\rho$ of a system of interest coupled to a reservoir (bath), generated by the Lindblad master equation (Liouvillian)
        \begin{align}
            i \frac{\partial}{\partial t}\rho = \mathcal{L}[\rho],
        \end{align}
        where the Lindblad superoperator is the generator of completely positive trace preserving (CPTP) time evolution and is given by \cite{lindblad1976generators,gorini1976completely}
        \begin{align}
            \mathcal{L} = [H,\rho] - i \frac{\Gamma}{2} \sum_m (\{J_m^\dagger J_m,\rho\}-2J_m\rho J_m^\dagger).
        \end{align}
        Here, $H$ is the system Hamiltonian and $J_m$ describe a set of quantum jump operators, weighted by a dissipation rate $\Gamma$ (with rate differences between different jump operators absorbed in the definition of $J_m$). The dissipative component describes time evolution under a non-Hermitian Hamiltonian $i\frac{\Gamma}{2} \sum_m J_m^\dag J_m$ subjected to quantum jumps to restore CPTP behavior---a key distinction from studies of classical non-Hermitian systems.

        Quadratic Lindbladians comprise a rich class of open quantum systems for which the Hamiltonian is bilinear in fermions/bosons
        \begin{align}
            H_{\bm{k}} &:= \sum_{\bm{k},\alpha,\beta} h_{\bm{k},\alpha\beta} c_{\bm{k},\alpha}^\dagger c_{\bm{k},\beta}\nonumber\\&\qquad +\sum_{\bm{k},\alpha,\beta} \Delta_{\bm{k},\alpha\beta} c_{\bm{k},\alpha}^\dagger c_{-\bm{k},\beta}^\dagger + \Delta_{\bm{k},\alpha\beta}^* c_{-\bm{k},\beta} c_{\bm{k},\alpha},
        \end{align}
        and jump operators are linear in fermions/bosons
        \begin{align}
            J_{\bm{k},m} := \sum_\alpha a_{\bm{k},m,\alpha} c_{\bm{k},\alpha} + b_{\bm{k},m,\alpha} c_{\bm{k},\alpha}^\dagger ~,
        \end{align}
        describing particle gain and loss processes. 
        This quadratic structure ensures that Wick's theorem can be used to obtain multi-point correlation functions in terms of two-point correlation functions. Additionally this quadratic structure leads to Gaussian path integrals with a single-particle matrix structure that can be leveraged to find these two-point functions: the single-particle Green's functions.

        The key object is the single-particle matrix $\Xi=\mathcal{H}+i\Sigma^R$, a combination of a Hermitian (coherent) term and an anti-Hermitian retarded self-energy contribution that describes Lindbladian dissipation. While $\Sigma^R$ originates from gain and loss processes due to coupling with a bath, we note that similar dissipative processes can emerge in interacting closed systems. Lindbladians describe frequency-independent $\Sigma^R$, which naturally arise from Markovian baths with broad spectral functions. The formalism developed in this paper is therefore applicable for steady state or meta-stable responses in any regimes that permit approximating the Keldysh self energy via frequency-independent Lindbladian self energies.
        
        \subsubsection{Single-Particle Fermionic Lindbladians}

            To illustrate the analogy between quadratic Lindbladians and quadratic Hamiltonians, consider defining ``left" and ``right" superfermions \cite{prosen2008quantum,prosen2010quantization}, which act on the left side ($\ell_\alpha \rho = c_\alpha\rho \mathcal{P}$) and right side ($r_\alpha \rho = \rho c_\alpha^\dag \mathcal{P}$) of the density matrix, with $\mathcal{P}$ the total fermionic parity operator. Here, $\alpha$ includes orbital/spin indices and continuous parameters other than the wave-vector. It is convenient to define these superoperators as ordinary operators acting on the vectorized density matrix, yielding
        	\begin{align}
        	   \ell_{\bm{k},\alpha} &:= c_{\bm{k},\alpha}\otimes \mathcal{P}\\
                \ell_{\bm{k},\alpha}^\dagger &:= c_{\bm{k},\alpha}^\dagger \otimes \mathcal{P}\\
                r_{\bm{k},\alpha} &:= \1 \otimes \mathcal{P} c_{\bm{k},\alpha}^\dagger\\
                r_{\bm{k},\alpha}^\dagger &:= \1 \otimes c_{\bm{k},\alpha} \mathcal{P},
        	\end{align}
            
            For finite-dimensional vector spaces, where matrix representations of the second quantized operator exist, the Lindblad master equation can be re-expressed in vectorized form as
            \begin{align}
                i\frac{\partial}{\partial t}\vec{\rho} = \hat{\mathcal{L}}\cdot\vec{\rho},
            \end{align}
            where $\vec{\rho}$ is a vector rather than a matrix and \cite{amshallem2015three}
        	\begin{align}
        	   \hat{\mathcal{L}} = &(\1\otimes H - H^*\otimes \1)\nonumber\\ &-i\frac{\Gamma}{2}\sum_m\left(\1\otimes J_m^\dagger J_m + (J_m^\dagger J_m)^*\otimes \1 - 2 J_m^*\otimes J_m\right),
        	\end{align}
        	which decomposes in terms of jump operators $J_{L/R}$ as \cite{talkington2022dissipation}
        	\begin{align}\label{eq:vectorized}
        	   \hat{\mathcal{L}} = &(H_{\bm{k},R}-H_{\bm{k},L})\nonumber\\ &- i\frac{\Gamma}{2}\sum_m \big(J_{\bm{k},m,R}^\dagger J_{\bm{k},m,R} + (J_{\bm{k},m,L}^\dagger J_{\bm{k},m,L})^*\nonumber\\&\qquad\quad\ \ - 2 J_{\bm{k},m,L}J_{\bm{k},m,R}^* \big),
        	\end{align}
            Here the operators are
        	\begin{align}
        	   H_{\bm{k},L} &:= \sum_{\alpha,\beta} h_{\bm{k},\alpha\beta} \ell_{\bm{k},\alpha}^\dagger \ell_{\bm{k},\beta}\nonumber\\&\qquad + \Delta_{\bm{k},\alpha\beta} \ell_{\bm{k},\alpha}^\dagger \ell_{-\bm{k},\beta}^\dagger + \Delta_{\bm{k,}\alpha\beta}^* \ell_{-\bm{k},\beta} \ell_{\bm{k},\alpha},\\
        	   H_{\bm{k},R} &:= \sum_{\alpha,\beta} h_{\bm{k},\alpha\beta} r_{\bm{k},\beta}^\dagger r_{\bm{k},\alpha} \nonumber\\&\qquad + \Delta_{\bm{k},\alpha\beta} r_{\bm{k},\alpha}^\dagger r_{-\bm{k},\beta}^\dagger + \Delta_{\bm{k},\alpha\beta}^* r_{-\bm{k},\beta} r_{\bm{k},\alpha}\\
        	   J_{\bm{k},m,L} &:= \sum_\alpha a_{\bm{k},m,\alpha}\ell_{\bm{k},\alpha} + b_{\bm{k},m,\alpha}\ell_{\bm{k},\alpha}^\dagger,\\
        	   J_{\bm{k},m,R} &:= \sum_{\alpha} - a_{\bm{k},m,\alpha}r_{\bm{k},\alpha} + b_{\bm{k},m,\alpha}r_{\bm{k},\alpha}^\dagger. 
        	\end{align}
            
            Expanding the products of jump operators above now defines three single-particle matrices that comprise the dissipative action
            \begin{align}\label{eq:matrices}
                A_{\bm{k}} &= \sum_m \bm{a}^*_{\bm{k},m} \bm{a}_{\bm{k},m}\\
                B_{\bm{k}} &= \sum_m \bm{b}_{-\bm{k},m} \bm{b}^*_{-\bm{k},m}\\
                C_{\bm{k}} &= \sum_m \bm{a}^*_{\bm{k},m} \bm{b}_{\bm{k},m},
            \end{align}
            where $\bm{a}_{\bm{k},m}=(a_{\bm{k},m,1},\dots,a_{\bm{k},m,N})$ and $\bm{b}_{\bm{k},m}=(b_{\bm{k},m,1},\dots,b_{\bm{k},m,N})$ are vectors of the jump operator coefficients and the matrices are constructed from the outer product of these coefficients.
            To make progress from here, we express $\hat{\mathcal{L}}$ in terms of a generalization of Prosen's ``third-quantization"  algebra \cite{prosen2008third,prosen2010spectral}.
            Defining $\bm{\ell} = (\ell_{1}, \ell_{2}, \dots, \ell_{N})$, and $\bm{r} = (r_{1}, r_{2}, \dots, r_{N})$ for $N$ orbitals, the Lindbladian superoperator can be written as
            \begin{align}\label{eq:Lquadratic}
                \hat{\mathcal{L}} = \bm{\Psi}_{\bm{k}}^\dag \cdot \left[ L^{\textrm{coh}}_{\bm{k}} - i L^{\textrm{dis}}_{\bm{k}} \right] \cdot \bm{\Psi}_{\bm{k}}
            \end{align}
            with ``third quantized" superfermions/superbosons $\Psi_{\bm{k}}^\dag = [ \bm{\ell}_{\bm{k}}^\dag, \bm{r}_{\bm{k}}^\dag, \bm{\ell}_{-\bm{k}}, \bm{r}_{-\bm{k}} ]$ \cite{prosen2008third,prosen2010spectral,talkington2022dissipation}. The single-particle matrices $L^{\textrm{coh}}_{\bm{k}}$ and $L^{\textrm{dis}}_{\bm{k}}$ are

            \begin{align}
                L^\text{coh}_{\bm{k}} =
                \begin{pmatrix}
                    H_{\bm{k}} & 0 & \Delta_{\bm{k}} & 0\\
                    0 & \pm H_{\bm{k}} & 0 & \mp \Delta_{\bm{k}}\\
                    \Delta^\dagger_{\bm{k}} & 0 & \mp H^\top_{-\bm{k}} & 0\\
                    0 & \mp\Delta^\dagger_{\bm{k}} & 0 & - H^\top_{-\bm{k}}
                \end{pmatrix}
            \end{align}
            and
            \begin{align}
                &L^\text{dis}_{\bm{k}} = \frac{\Gamma}{2}\times\nonumber\\
                &\begin{pmatrix}
                    A_{\bm{k}}\mp B_{\bm{k}}&-2B_{\bm{k}}&C_{\bm{k}}\mp C^\top_{-\bm{k}}&2C^\top_{-\bm{k}}\\
                    -2A_{\bm{k}}&\mp A_{\bm{k}}+B_{\bm{k}}&-2C_{\bm{k}}&\pm C_{\bm{k}}-C^\top_{-\bm{k}}\\
                    C^\dagger_{\bm{k}} \mp C^*_{-\bm{k}}&-2C^*_{-\bm{k}}&\mp A^\top_{-\bm{k}} + B^\top_{-\bm{k}} & 2A^\top_{-\bm{k}}\\
                    2C^\dagger_{\bm{k}}&\pm C^\dagger_{\bm{k}} - C^*_{-\bm{k}}&2B^\top_{-\bm{k}}& A^\top_{-\bm{k}}\mp B^\top_{-\bm{k}}
                \end{pmatrix},
            \end{align}
            where the top and bottom of $\pm$ and $\mp$ refer to fermions and bosons respectively.
            
            Here we note that the key to writing the Keldysh functional integral in the vectorized space is to reexpress operators in terms of their eigenvalues upon the application of complete sets of coherent states as detailed in \cite{sieberer2016keldysh,kamenev2023field}; in this formalism the parity operators are naturally absorbed into the eigenvalues.
            
        \subsubsection{Single-Particle Fermionic Green's Functions}\label{sec:action}

            \begin{figure}
                \centering
                \includegraphics[width=0.6\linewidth]{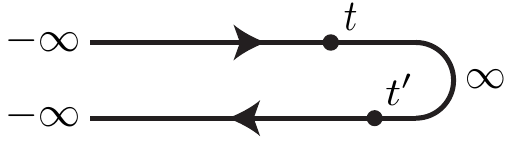}
                \caption{Schematic illustration of the Keldysh contour over which the evolution of the density matrix is conducted. Right fermions $r$ live on the right-moving contour. Left fermions $\ell$ live on the left-moving contour.}
                \label{fig:contour}
            \end{figure}
        
            To compute Green's functions, we will be interested in the generating functional \cite{sieberer2016keldysh}
            \begin{align}\label{eq:partition}
                Z = \int \mathcal{D}[\bm{\ell},\bm{r},\bm{\ell}^\dagger,\bm{r}^\dagger]\ e^{iS},
            \end{align}
            where vectors of left and right-contour fermion/boson fields are given by $\bm{\ell} = (\ell_{1}, \ell_{2}, \dots, \ell_{N})$, and $\bm{r} = (r_{1}, r_{2}, \dots, r_{N})$ for $N$ orbitals. Here we use $\ell^\dagger$, $r^\dagger$ to indicate the Grassman conjugate field of $\ell$, $r$. The left fermions correspond to operators acting on the left/contour moving to the left, and the right fermions correspond to operators acting on the right/contour moving to the right as illustrated in Fig. \ref{fig:contour}.
            The continuous-time Lindblad-Keldysh action $S$ is formally given by \cite{sieberer2016keldysh}
            \begin{align}\label{eq:action}
                S &= \int_{-\infty}^\infty dt \int_\mathrm{BZ}\frac{d^D\bm{k}}{(2\pi)^D}\ \left(\bm{\ell}_{\bm{k}}^\dagger i\partial_t\1 \bm{\ell}_{\bm{k}} - \bm{r}_{\bm{k}}^\dagger i\partial_t\1 \bm{r}_{\bm{k}}\right) - \hat{\mathcal{L}}_{\bm{k}},
            \end{align}
            where $\bm{k}$ is the wave-vector.
            
            For quadratic Lindbladians, the expression for $\hat{\mathcal{L}}$ given in Eq. (\ref{eq:vectorized}) can be replaced by the quadratic superoperator of Eq. (\ref{eq:Lquadratic}) written using left/right propagating fields and one obtains
            \begin{align}\label{eq:action-sp}
                S = \int\limits_{-\infty}^\infty dt\int_\mathrm{BZ}\frac{d^D\bm{k}}{(2\pi)^D} \ \bm{\Psi}_{\bm{k}}^\dagger
                \bigg[
                    i\partial_t\bm{1} - L^\text{coh}_{\bm{k}} + i L^\text{dis}_{\bm{k}}
                \bigg]
                \bm{\Psi}_{\bm{k}}.
            \end{align}
            Now, the physical responses are more transparent in a rotated basis obtained using the Keldysh-Larkin-Ovchinnikov transformation \cite{rammer1986quantum,larkin1986}. For fermions the rotation matrix is
            \begin{align}
            U_L = \frac{1}{\sqrt{2}}
            \begin{pmatrix}
                1&-1&0&0\\
                0&0&1&1\\
                1&1&0&0\\
                0&0&1&-1
            \end{pmatrix},
            \end{align}
            so that $\widetilde{\bm{\Psi}} =(\bm{\psi}_1,\bm{\psi}_2,\bar{\bm{\psi}}_1,\bar{\bm{\psi}}_2):=U_L (\bm{\ell},\bm{r},\bm{\ell}^\dagger,\bm{r}^\dagger)=U_L\bm{\Psi}$. Note that $\bm{\psi}_i$ and $\bar{\bm{\psi}}_i$ are independent fields and while they are adjoints for bosons, they are unrelated for fermions \cite{kamenev2023field}.     
            In this new basis,
            \begin{align}
                S = \int_{-\infty}^\infty dt\int_\mathrm{BZ}\frac{d^D\bm{k}}{(2\pi)^D} \ \bar{\widetilde{\bm{\Psi}}}_{\bm{k}}
                \bigg[
                    i\partial_t\bm{1} - \widetilde{L}^\text{coh}_{\bm{k}} + i \widetilde{L}^\text{dis}_{\bm{k}}
                \bigg]
                \widetilde{\bm{\Psi}}_{\bm{k}}
            \end{align}
            for transformed single-particle matrices \cite{talkington2022dissipation}
            \begin{align}
                \widetilde{L}^\text{coh}_{\bm{k}} =
                \begin{pmatrix}
                    h_{\bm{k}} & \Delta_{\bm{k}} & 0 & 0\\
                    \Delta^\dagger_{\bm{k}} & -h^\top_{-\bm{k}} & 0 & 0\\
                    0 & 0 & h_{\bm{k}} & \Delta_{\bm{k}}\\
                    0 & 0 & \Delta^\dagger_{\bm{k}} & - h^\top_{-\bm{k}}
                \end{pmatrix}
            \end{align}
            and
            \begingroup
            \setlength{\arraycolsep}{0 pt}
            \begin{align}
                &\widetilde{L}^\text{dis}_{\bm{k}} = \frac{\Gamma}{2} \times\nonumber\\
                &\begin{pmatrix}
                    A_{\bm{k}}+B_{\bm{k}} & C_{\bm{k}}+C^\top_{-\bm{k}} & 2(A_{\bm{k}}-B_{\bm{k}}) & 2(C_{\bm{k}}-C^\top_{-\bm{k}})\\
                    C^\dagger_{\bm{k}} + C^*_{-\bm{k}} & A^\top_{-\bm{k}} + B^\top_{-\bm{k}} & 2(C^\dagger_{\bm{k}}-C^*_{-\bm{k}}) & -2(A_{\bm{k}}-B_{\bm{k}})\\
                    0&0&-(A_{\bm{k}}+B_{\bm{k}})&-(C_{\bm{k}}+C^\top_{-\bm{k}})\\
                    0&0&-(C^\dagger_{\bm{k}}+C^*_{-\bm{k}})&-(A^\top_{-\bm{k}}+B^\top_{-\bm{k}})
                \end{pmatrix},
            \end{align}
            \endgroup
            These last expressions have recently also been obtained by a field-theoretic treatment conducted by Thompson and Kamenev in Ref. \cite{thompson2023field}, and are equivalent to the BdG form we obtained using third quantization in Ref. \cite{talkington2022dissipation} and the form obtained by McDonald and Clerk in \cite{mcdonald2023third}.
            In this rotated basis one arrives at the inverse of the single particle Keldysh Green's functions \cite{kamenev2023field}
            \begin{align}
                \mathcal{G}^{-1}_{\bm{k}} = i\partial_t\bm{1} - \widetilde{L}^\text{coh}_{\bm{k}} + i \widetilde{L}^\text{dis}_{\bm{k}}  \label{eq:GkeldyshBlock}
            \end{align}
            for Lindbladians, which define spectroscopic response properties discussed below and take the usual Keldysh block structure
            \begin{align}
                \mathcal{G}_{\bm{k}} = \begin{pmatrix}
                    G^R_{\bm{k}} & G^K_{\bm{k}}\\
                    0 & G^A_{\bm{k}}
                \end{pmatrix}.
            \end{align}
            The Green's functions can be succinctly written by introducing an effective \textit{non-Hermitian} Hamiltonian
            \begin{align}
                \Xi_{\bm{k}} &=
                \begin{pmatrix}
                    h_{\bm{k}} & \Delta_{\bm{k}}\\
                    \Delta^\dagger_{\bm{k}} & - h^\top_{-\bm{k}}
                \end{pmatrix}
                - i \frac{\Gamma}{2}
                \begin{pmatrix}
                    A_{\bm{k}}+B_{\bm{k}} & C_{\bm{k}}+C^\top_{-\bm{k}}\\
                    C^\dagger_{\bm{k}}+C^*_{-\bm{k}} & A^\top_{-\bm{k}} + B^\top_{-\bm{k}}
                \end{pmatrix} \notag\\
                & \equiv \mathcal{H}_{\bm{k}} + i \Sigma^R_{\bm{k}} ,
            \end{align}
            with $\Sigma^R$ the retarded self energy, and the Lindblad-Keldysh self energy
            \begin{align}
                \Sigma^K_{\bm{k}} = -i \Gamma
                \begin{pmatrix}
                    A_{\bm{k}}-B_{\bm{k}} & C_{\bm{k}}-C^\top_{-\bm{k}}\\
                    C^\dagger_{\bm{k}}-C^*_{-\bm{k}} & -A^\top_{-\bm{k}} + B^\top_{-\bm{k}}
                \end{pmatrix},
            \end{align}
            Inverting Eq. (\ref{eq:GkeldyshBlock}), one obtains retarded and advanced Green's functions
            \begin{align}
                G^R_{\bm{k}} &= [i\partial_t \bm{1} - \Xi_{\bm{k}}]^{-1}\\
                G^A_{\bm{k}} &= [i\partial_t \bm{1} - \Xi^\dagger_{\bm{k}}]^{-1},
            \end{align}
            where $G^A = (G^R)^\dagger$, since $H=H^\dagger$, $A=A^\dagger$, and $B=B^\dagger$. The retarded/advanced Green's functions encode the effective non-Hermitian single-particle spectrum that dictates spectroscopic responses. Conversely, the Keldysh Green's function can be obtained via
            \begin{align}
                G^K_{\bm{k}} = G^R_{\bm{k}}\Sigma^K_{\bm{k}} G^A_{\bm{k}}
            \end{align}
            which is true, only in the case of the steady state. For transient behavior $\Sigma^K$ takes a different form and $G^K=G^R F - F G^A$ for some Hermitian matrix $F$ that specified the distribution function \cite{kamenev2023field,thompson2023field}.
            Analogously to in closed systems one may consider the Fourier transform of these expressions to frequency space where $i\partial_t \mapsto \omega$ and we do not need to include an \textit{infinitesimal} $i\eta$ since there is naturally a \textit{finite} self energy in these open systems. Doing so we obtain
            \begin{align}
                G^R_{\bm{k}} &= [\omega \bm{1} - \Xi_{\bm{k}}]^{-1}\\
                G^A_{\bm{k}} &= [\omega \bm{1} - \Xi^\dagger_{\bm{k}}]^{-1},
            \end{align}
            which can be decomposed in terms of the right $|u\rangle$ and left $\langle \bar{u}|$ eigenstates of $\Xi$ as 
            \begin{align}
                G^R_{\bm{k}}(\omega) &= \sum_{n} \frac{|u_{\bm{k},n}\rangle\langle\bar{u}_{\bm{k},n}|}{\omega - \xi_{\bm{k},n}},
            \end{align}
            and
            \begin{align}
                G^A_{\bm{k}}(\omega) &= \sum_n \frac{|\bar{u}_{\bm{k},n}\rangle \langle u_{\bm{k},n}|}{\omega-\xi_{\bm{k},n}^*}.
            \end{align}

        \subsubsection{Quadratic Bosonic Lindbladians}

            For bosonic systems, the generating functional Eq. (\ref{eq:partition}) and action Eq. (\ref{eq:action}-\ref{eq:action-sp}) are the exact same and analogous arguments define ``left" and ``right" superbosons
        	\begin{align}
            	\ell_{\bm{k},\alpha} := c_{\bm{k},\alpha}\otimes \1\\
                \ell_{\bm{k},\alpha}^\dagger := c_{\bm{k},\alpha}^\dagger \otimes \1\\
                r_{\bm{k},\alpha}:= \1 \otimes c_{\bm{k},\alpha}^\dagger\\
                r_{\bm{k},\alpha}^\dagger:= \1 \otimes c_{\bm{k},\alpha},
        	\end{align}
            where we note that Eq. (\ref{eq:vectorized}) has the same form except that $J_{m,R} := \sum_{\alpha} a_{m,\alpha}r_\alpha + b_{m,\alpha}r_\alpha^\dagger$ for bosons.
            The $A$, $B$, and $C$ matrices have the same form as in the fermionic case and are given by Eq. (\ref{eq:matrices}).
            
            Now we can perform the Keldysh-Larkin-Ovchinnikov rotation \cite{rammer1986quantum}, where for bosons the rotation matrix is
            \begin{align}
                U_L = \frac{1}{\sqrt{2}}
                \begin{pmatrix}
                    1&1&0&0\\
                    0&0&1&-1\\
                    1&-1&0&0\\
                    0&0&1&1
                \end{pmatrix},
            \end{align}
            so that $\widetilde{\bm{\Psi}} = (\bm{\psi}_1,\bm{\psi}_2,\bar{\bm{\psi}}_1,\bar{\bm{\psi}}_2):=U_L (\bm{\ell},\bm{r},\bm{\ell}^\dagger,\bm{r}^\dagger)=U_L\bm{\Psi}$.
            
            Expressed in the new basis, we have
            \begin{align}
                \widetilde{L}^\text{coh}_{\bm{k}} =
                \begin{pmatrix}
                    0 & 0 & h_{\bm{k}} & \Delta_{\bm{k}}\\
                    0 & 0 & \Delta^\dagger_{\bm{k}} & h^\top_{-\bm{k}}\\
                    h_{\bm{k}} & \Delta_{\bm{k}} & 0 & 0\\
                    \Delta^\dagger_{\bm{k}} & h^\top_{-\bm{k}} & 0 & 0
                \end{pmatrix}
            \end{align}
            and
            \begingroup
            \setlength{\arraycolsep}{2 pt}
            \begin{align}
                &\widetilde{L}^\text{dis}_{\bm{k}} = \frac{\Gamma}{2} \times \nonumber \\ &
                \begin{pmatrix}
                    0&0&-(A_{\bm{k}}-B_{\bm{k}})&-(C_{\bm{k}}-C^\top_{-\bm{k}})\\
                    0&0&-(C^\dagger_{\bm{k}}-C^*_{-\bm{k}})&A^\top_{-\bm{k}}-B^\top_{-\bm{k}}\\
                    A_{\bm{k}}-B_{\bm{k}}&C_{\bm{k}}-C^\top_{-\bm{k}} & 2(A_{\bm{k}}+B_{\bm{k}}) & 2(C_{\bm{k}}+C^\top_{-\bm{k}})\\
                    C^\dagger_{\bm{k}}-C^*_{-\bm{k}} & B^\top_{-\bm{k}} -A^\top_{-\bm{k}} & 2(C^\dagger_{\bm{k}}+C^*_{-\bm{k}}) & 2(A^\top_{-\bm{k}}+B^\top_{-\bm{k}})
                \end{pmatrix}\!,
            \end{align}
            \endgroup
            which will be key to obtaining the Green’s functions and calculating response properties. 

            For bosons, the Keldysh block Green's function takes the block structure
            \begin{align}
                \mathcal{G}_{\bm{k}} =
                \begin{pmatrix}
                    G^K_{\bm{k}} & G^R_{\bm{k}}\\
                    G^A_{\bm{k}} & 0
                \end{pmatrix}.
            \end{align}
            Introducing again an effective non-Hermitian Hamiltonian $\Xi=\mathcal{H}+i \Sigma^R$ for bosons
            \begin{align}
                \Xi_{\bm{k}} =
                \begin{pmatrix}
                    h_{\bm{k}} & \Delta_{\bm{k}} \\
                    \Delta^\dagger_{\bm{k}} & h^\top_{-\bm{k}}
                \end{pmatrix}
                - i \frac{\Gamma}{2}
                \begin{pmatrix}
                    A_{\bm{k}}-B_{\bm{k}} & C_{\bm{k}}-C^\top_{-\bm{k}}\\
                    C^\dagger_{\bm{k}}-C^*_{-\bm{k}} & B^\top_{-\bm{k}} - A^\top_{-\bm{k}}
                \end{pmatrix},
            \end{align}
            where contrary to the fermionic case the imaginary part of $\Sigma^R$ can be negative corresponding to possible gain processes that are impossible in the fermionic case. This means that contrary to a ``fully filled" steady state in fermionic systems there are bosonic systems that have physical choices for the dissipative jump operators where there is no steady state due to unbounded gain. Generically, the precise condition for there to be no steady state is that $\sigma^3\Xi$ (or if $C=\Delta=0$ then $h-i\frac{\Gamma}{2}(A-B)$) acquires an eigenvalue with positive imaginary part: in this case there will be a mode that grows exponentially with time. Expressions for densities and correlation functions below assume that all poles are in the negative half-plane.
            The retarded and advanced Green's functions are
            \begin{align}
                G^R_{\bm{k}} &= [i\partial_t \sigma^3 - \Xi_{\bm{k}}]^{-1}\\
                G^A_{\bm{k}} &= [i\partial_t \sigma^3 -  \Xi^\dagger_{\bm{k}}]^{-1}
            \end{align}
            with $G^A = (G^R)^\dagger$. The Pauli $\sigma^3$ matrix acting in Nambu space (and as the identity in orbital space) arises since $\bar{\psi}$ and $\psi$ fields are related for bosons (but are independent for fermions).
            The Keldysh self energy term reads
            \begin{align}\label{eq:keldysh-sigma}
                \Sigma^K_{\bm{k}} = -i \frac{\Gamma}{2}
                \begin{pmatrix}
                    A_{\bm{k}}+B_{\bm{k}} & C_{\bm{k}}+C^\top_{-\bm{k}}\\
                    C^\dagger_{\bm{k}}+C^*_{-\bm{k}} & A^\top_{-\bm{k}} + B^\top_{-\bm{k}}
                \end{pmatrix},
            \end{align}
            and the Keldysh component of the Green's function again becomes $G^K = G^R\Sigma^K G^A$ for the steady state response.

        \subsubsection{Spectral Representation and Non-Hermitian Single-Particle Band Structure}
    
            Importantly, spectroscopic responses for quadratic Lindbladians generalize the response formalism of band insulators and metals in a two-fold manner: first, the effective single-particle band structure generalizes to a non-Hermitian problem, with the disparity of left and right eigenvectors having important consequences on, e.g., optical transition matrix elements. Second, the steady-state distribution of dissipative systems is generically distinct from thermal distributions in equilibrium.
            
            To discuss ramifications of the non-Hermitian spectrum that enters into the definition of retarded and advanced Green's functions, we start by describing its eigenmodes. We have right and left eigenvectors
            \begin{align}
                (\sigma^i \Xi_{\bm{k}}) |u_{{\bm{k}},n}\rangle &= \xi_{{\bm{k}},n} |u_{{\bm{k}},n}\rangle\\
                \langle\bar{u}_{{\bm{k}},n}| (\sigma^i\Xi_{\bm{k}}) &= \xi_{{\bm{k}},n} \langle\bar{u}_{{\bm{k}},n}|,
            \end{align}
            where $\sigma^i=\1$ for fermions and $\sigma^i=\sigma^3$ for bosons which acts on the Nambu structure as $\sigma^3$ and as the identity on the orbital structure.
            Here, $\bra{u_n} = \ket{u_n}^\dag \neq \bra{\bar{u}_n}$, and
            \begin{align}
                \bra{u_{\bm{k},n}} (\sigma^i\Xi_{\bm{k}})^\dag = \xi_{{\bm{k}},n}^* \bra{u_{{\bm{k}},n}} \\
                (\sigma^i\Xi_{\bm{k}})^\dag \ket{\bar{u}_{{\bm{k}},n}} = \xi_{{\bm{k}},n}^* \ket{\bar{u}_{{\bm{k}},n}}.
            \end{align}
            Left and right eigenvectors are biorthogonal for all $\bm{k}$
            \begin{align}
                \langle\bar{u}_{{\bm{k}},n}| u_{{\bm{k}},n'}\rangle = \delta_{nn'},
            \end{align}
            and completeness (away from exceptional points) \cite{hashemi2022linear} dictates
            \begin{align}
                \bm{1} & = \sum_n |u_{{\bm{k}},n}\rangle\langle \bar{u}_{{\bm{k}},n}|\\
                \sigma^i \Xi_{\bm{k}} &= \sum_n \xi_{{\bm{k}},n} |u_{{\bm{k}},n}\rangle\langle \bar{u}_{{\bm{k}},n}|.
            \end{align}
            This permits a spectral representation of the steady-state single particle Green's functions
            \begin{align}
                G^R_{\bm{k}}(\omega) &= \sum_{n} \frac{|u_{{\bm{k}},n}\rangle\langle\bar{u}_{{\bm{k}},n}|}{\omega - \xi_{\bm{k},n}} \sigma^i, \\
                G^A_{\bm{k}}(\omega) &= \sum_n \sigma^i \frac{|\bar{u}_{{\bm{k}},n}\rangle \langle u_{{\bm{k}},n}|}{\omega-\xi_{{\bm{k}},n}^*},
            \end{align}
            Finally, for the non-equilibrium steady-state, the Keldysh Green's function can be written as
            \begin{align}\label{eq:GK-ss}
                G^K_{\bm{k}}(\omega) &= G^R_{\bm{k}}\Sigma^K_{\bm{k}} G_{\bm{k}}^A\\
                &= \sum_{n,n'} \frac{\langle\bar{u}_{{\bm{k}},n}|\sigma^i\Sigma^K_{\bm{k}}\sigma^i|\bar{u}_{{\bm{k}},n'}\rangle}{(\omega - \xi_{{\bm{k}},n})(\omega-\xi_{{\bm{k}},n'}^*)} |u_{{\bm{k}},n}\rangle\langle u_{{\bm{k}},n'}|.
            \end{align}
        
        \subsubsection{Green's Functions}

            We will be interested in evaluating two-point correlation functions which follow naturally by considering the lesser and greater Green's functions
            \begin{align}
                G^<_{\alpha\beta}(t,t') &= - i e^{i\phi} \langle c_\beta^\dagger(t') c_\alpha(t)\rangle\\
                G^>_{\alpha\beta}(t,t') &= -i\langle c_\alpha(t) c_\beta^\dagger(t')\rangle,
            \end{align}
            where $\phi=\pi$ for fermions and $2\pi$ for bosons is the exchange angle.
            Expressed in terms of $G^<$ and $G^>$ the retarded, advanced, and Keldysh Green's functions become \cite{rammer1986quantum}
            \begin{align}\label{eq:Grak}
                G^R_{\alpha\beta}(t,t') &= +\theta(t-t')[G_{\alpha\beta}^>(t,t')-G_{\alpha\beta}^<(t,t')]\\
                G^A_{\alpha\beta}(t,t') &= -\theta(t'-t)[G_{\alpha\beta}^>(t,t')-G_{\alpha\beta}^<(t,t')]\\
                G^K_{\alpha\beta}(t,t') &= G_{\alpha\beta}^>(t,t') + G_{\alpha\beta}^<(t,t').
            \end{align}
            This naturally leads to the identity
            \begin{align}
                G^R - G^A = G^> - G^<,
            \end{align}
            The two point correlation functions read
            \begin{align}\label{eq:two-point}
                \langle c_\alpha^\dagger(t) c_\beta(t')\rangle &= \frac{i}{2} e^{i\phi}(G^K - G^R + G^A)_{\beta\alpha}(t',t)\\
                \langle c_\alpha(t) c_\beta^\dagger(t')\rangle &= \frac{i}{2}(G^K + G^R - G^A)_{\alpha\beta}(t,t'),
            \end{align}
            where we used that $1/e^{i\phi}=e^{i\phi}$ for fermions and bosons, and reindexed in the first line. For quadratic Lindbladians, multi-point correlation functions can now be computed straightforwardly using Wick's theorem.

    \subsection{Spectroscopic Response Formalism} \label{sec:response}

        Armed with a representation of Lindblad-Keldysh Green's functions in terms of effective non-Hermitian spectra, we now transcribe linear and non-linear dynamical responses for dissipative system into the framework of quadratic Lindbladians. Detailed derivations of these frequency dependent response functions is given in the Supplemental Material \footnote{See Supplemental Material, Section 2 and 3 for derivations of the frequency dependent response functions.}.

        \subsubsection{Spectral Function}

            To set the stage, we start by considering the single-particle spectral function that describes physical single-particle excitations and their density of states. The spectral function is given by
            \begin{align}\label{eq:spectral}
                A_{\bm{k}}(\omega) &= - \frac{1}{\pi}\, \text{Im}(\text{Tr}[G_{\bm{k}}^R(\omega)])\\
                &= - \frac{1}{\pi} \text{Im}(\text{Tr}[\sum_n \frac{|u_{{\bm{k}},n}\rangle\langle\bar{u}_{{\bm{k}},n}|}{\omega-\xi_{{\bm{k}},n}}\sigma^i])
            \end{align}
            where $\sigma^i=\1$ for fermions and $\sigma^i=\sigma^3$ which acts on the Nambu structure for bosons.
            This can then be used to obtain the density of states
            \begin{align}
                \rho(\omega) = \int_\text{BZ} \frac{d^D\bm{k}}{(2\pi)^D}\ A_{\bm{k}}(\omega),
            \end{align}
            and the number of bands is then given by
            \begin{align}
                N_{\bm{k}} = \int_{-\infty}^\infty d\omega\ A_{\bm{k}}(\omega),
            \end{align}
            where $N_{\bm{k}}$ is the same for all $\bm{k}$.

        \subsubsection{Particle Density and Equal-Time Expectation Values}\label{sec:density}

            To obtain the steady-state particle density, consider the equal time expectation value
            \begin{align}
                \langle c_{\bm{k},\alpha}^\dagger(t) c_{\bm{k},\beta}(t)\rangle &= \frac{i}{2} e^{i\phi}(G^K_{\bm{k},\beta\alpha} - G^R_{\bm{k},\beta\alpha} + G^A_{\bm{k},\beta\alpha})(t,t)\\
                &= \frac{1}{2}e^{i\phi}\left(\delta_{\beta\alpha} - iG_{\bm{k},\beta\alpha}^K(t,t)\right),
            \end{align}
            which follows from what we obtained above in Eq. (\ref{eq:two-point}). Using Eq. (\ref{eq:GK-ss}), one immediately finds the density using the frequency-space Keldysh Green's function; we obtain
            \begin{align}\label{eq:density}
                \langle c_{\bm{k},\alpha}^\dagger(t) &c_{\bm{k},\beta}(t)\rangle
                = \frac{e^{i\phi}}{2}\bigg(\bm{1}\nonumber\\& + \sum_{n,n'} \frac{\langle\bar{u}_{{\bm{k}},n} |\sigma^i \Sigma^K_{{\bm{k}}}\sigma^i |\bar{u}_{{\bm{k}},n'}\rangle}{\xi_{{\bm{k}},n'}^* - \xi_{{\bm{k}},n}} |u_{{\bm{k}},n}\rangle\langle u_{{\bm{k}},n'}|\bigg),
            \end{align}
            where we used the residue theorem to complete the $\omega$ integral and $\sigma^i=\1$ for fermions and $\sigma^i=\sigma^3$ which acts on the Nambu structure for bosons. The density is then the diagonal elements $\langle n_\alpha \rangle = \langle c_\alpha^\dagger c_\alpha\rangle$.

                By analogy, the steady state expectation value for an arbitrary single-particle operator
                \begin{align}\label{eq:single-particle-op}
                    \mathsf{O} = \sum_{\bm{k},\alpha,\beta} O_{\bm{k},\alpha\beta} c_{\bm{k},\alpha}^\dagger c_{\bm{k},\beta},
                \end{align}
                can be computed from the dissipative eigenbasis
                \begin{align}\label{eq:dia_response}
                \langle \mathsf{O} \rangle &= \frac{e^{i\phi}}{2}\bigg(\text{Tr}[O_{\bm{k}}]\nonumber\\ &\quad+ \sum_{\bm{k}nn'} \frac{\langle\bar{u}_{{\bm{k}},n} |\sigma^i\Sigma^K_{{\bm{k}}}\sigma^i|\bar{u}_{{\bm{k}},n'}\rangle}{\xi_{{\bm{k}},n'}^* - \xi_{{\bm{k}},n}} \langle u_{{\bm{k}},n'}| O_{\bm{k}} |u_{{\bm{k}},n}\rangle\bigg),
                \end{align}
                where $O_{\bm{k}}$ is the $2N\times 2N$ matrix with entries $[O_{\bm{k}}]_{\alpha\beta}=O_{\bm{k},\alpha\beta}$ and the trace runs over both state indices $n$ and momentum $\bm{k}$. Note that $O_{\bm{k}}$ is typically traceless in Nambu form, if time-reversal symmetry is respected.

        \subsubsection{Linear Response}

\begin{figure}
                \centering
                \includegraphics[width=\linewidth]{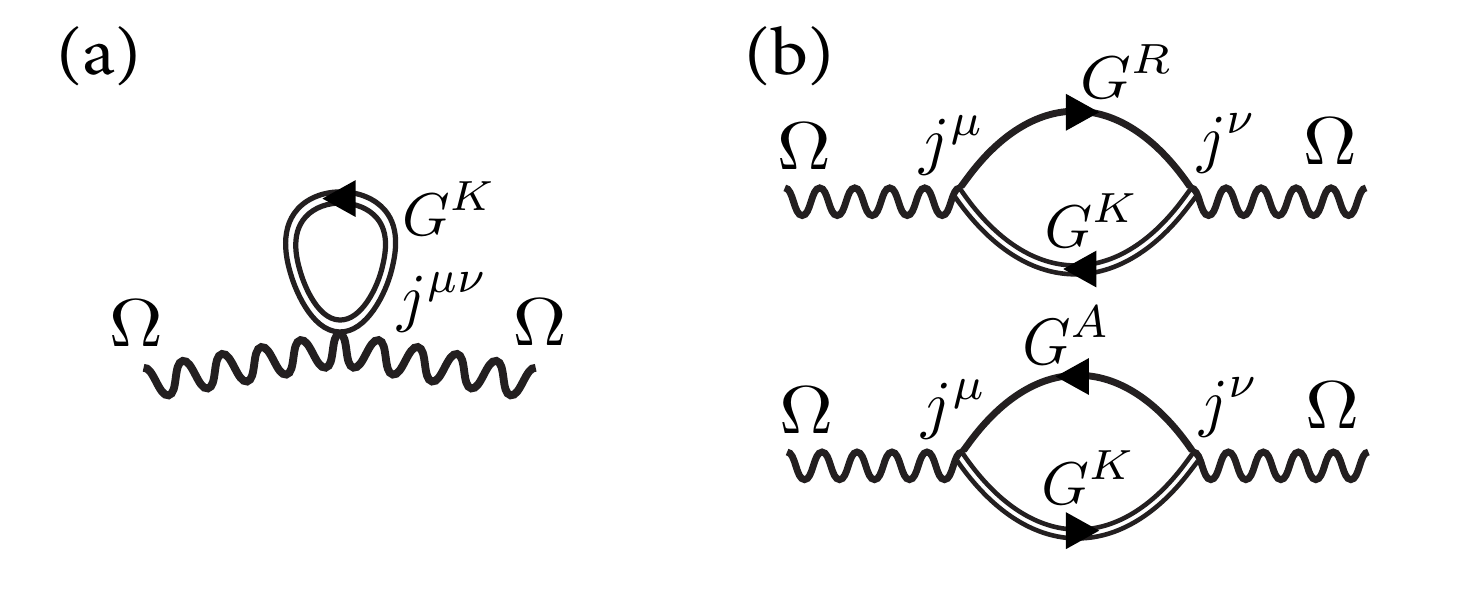}
                \caption{Feynman diagrams for dissipative linear response. (a) The density/diamagnetic response at frequency $\Omega$ is given by a trace over the Keldysh Green's function and vertex $O$ here taken as a current vertex for electromagnetic response $O=j^{\mu\nu}$. (b) The paramagnetic response is composed two diagrams involving Keldysh, retarded, and advanced Green's functions,  illustrated for the case where $O=j^\mu$ and $O'=j^\nu$ to describe electromagnetic response.}
                \label{fig:linear-feynman}
            \end{figure}

            We now turn to the steady state linear response of an arbitrary operator $\mathsf{O}(t)$ to a time-dependent perturbation $\mathsf{O}'(t)$, given by the correlation function
            \begin{align}
                    \Pi(\Omega) &= -i \int dt ~e^{-i\Omega t} \langle [\mathsf{O}(t),\mathsf{O}'(0)]\rangle \theta(t),
            \end{align}
            Generically, a Kubo formula can also be derived for dissipative systems. This is presented in the Supplemental Material \footnote{See Supplemental Material, Section 1 for a derivation of the Kubo formula for Lindbladian systems.} and remains generically applicable for interacting Lindbladians.
            For quadratic Lindbladians, Wick's theorem permits expressing the correlation function in terms of single-particle Keldysh Green's functions in frequency space
                \begin{align}\label{eq:paramagnetic}
                    \Pi(\Omega) &= \frac{i}{2} e^{i\phi}\int_{-\infty}^\infty \frac{d\omega}{2\pi}\ \bigg(\text{Tr}[O_{\bm{k}} G^R_{\bm{k}}(\omega)O'_{\bm{k}} G^K_{\bm{k}}(\omega+\Omega)]\nonumber\\&\hspace{0.9 in}+\text{Tr}[O_{\bm{k}} G^K_{\bm{k}}(\omega-\Omega)O'_{\bm{k}} G^A_{\bm{k}}(\omega)]\bigg),
                \end{align}
                where the details are given in the Supplemental Material \footnote{See Supplemental Material, Section 2 for derivations of the frequency dependent linear response functions.}. Expressing in terms of a spectral representation and completing the integral over frequencies $\omega$ using the residue theorem we find
                \begin{widetext}
                \begin{align}\label{eq:para_response}
                \Pi(\Omega)
                = -e^{i\phi}\!\sum_{{\bm{k}},n,n'} \!\! \frac{\langle\bar{u}_{\bm{k},n}|\sigma^i\Sigma^K_{\bm{k}}\sigma^i|\bar{u}_{\bm{k},n'}\rangle}{\xi_{\bm{k},n'}^*-\xi_{\bm{k},n}}
                \bigg(\sum_{n''}\frac{\langle u_{\bm{k},n'}|O_{\bm{k}}|u_{\bm{k},n''}\rangle\langle\bar{u}_{\bm{k},n''}|\sigma^iO'_{\bm{k}}|u_{\bm{k},n}\rangle}{(\xi_{\bm{k},n'}^*-\xi_{\bm{k},n''})-\Omega} + \frac{\langle u_{\bm{k},n'}|O'_{\bm{k}}\sigma^i|\bar{u}_{\bm{k},n''}\rangle\langle u_{\bm{k},n''}|O_{\bm{k}}|u_{\bm{k},n}\rangle}{(\xi_{\bm{k},n}-\xi_{\bm{k},n''}^*)+\Omega}\bigg)
                \end{align}
                \end{widetext}
                where $\phi=\pi$ and $\sigma^i=\1$ for fermions and $\phi=2\pi$ and $\sigma^i=\sigma^3$ which acts on the Nambu structure for bosons.
                The response corresponds diagrammatically to two loops as illustrated in Fig. \ref{fig:linear-feynman}, and obeys standard Kramers-Kronig relations between real and imaginary parts.
    
            \subsubsubsection{Example: Optical Conductivity and $f$-Sum Rule}

                Consider now the steady state linear optical response of a dissipative electron systems. The linear optical conductivity in velocity gauge is given by
                \begin{align}
                \sigma^{\mu\nu}(\Omega) = \frac{i}{\Omega}\Pi_\text{dia}^{\mu\nu} + \frac{i}{\Omega}\Pi_\text{para}^{\mu\nu}(\Omega)
                \end{align}
                where $\Pi_{\text{para}}$ is given by Eq. (\ref{eq:paramagnetic}) with $O_{\bm{k}} = j^{\mu}$ the paramagnetic current operator, and $\Pi_\text{dia}^{\mu\nu}=-i\langle j^{\mu\nu}\rangle$ with $j^{\mu\nu}$ the diamagnetic current operator. 
                
                In an open quantum system, a definition of a \textit{local} current operator within the system is given by the time derivative of the electric dipole operator $e \bm{x} = ie\, \nabla_k$ \cite{blount1962formalisms}, which in a Lindbladian setting yields an equation of motion $j^\mu = -i e \bar{\mathcal{L}}\{ x^\mu \}$ with $\bar{\mathcal{L}}$ the adjoint Lindbladian for the Heisenberg equation of motion. We take $e=1$ in the discussion that follows. For simplicity we will consider situations where the bath cannot carry a current, leading to momentum-independent jump operators. In this situation, the current operator recovers the usual closed-system form $j^\mu=i[\mathcal{H},x^\mu]$ with $\mathcal{H}$ the coherent system Hamiltonian.

                In closed systems, the frequency-sum ($f$-sum) rule for optical responses \cite{mahan2000many} relates the integral over the paramagnetic (real part) of the optical response to the ground state expectation value of the diamagnetic current
                \begin{align}
                \int d\Omega\ \sigma^{xx}_\text{para}(\Omega) = 2\pi \langle j^{xx}_\text{dia} \rangle ~,
                \end{align}
                a relation which is satisfied at \textit{each} $\bm{k}$-point individually and permits counting the number of carriers by measuring the frequency-dependent absorption. In quadratic Lindbladian systems we find that a $f$-sum rule holds for the $\bm{k}$ \textit{integrated} response
                for a bath that does not contribute to the current (with Lindbladian jump operators that are momentum-independent, i.e. $\bm{a}$ and $\bm{b}$ are momentum independent). Proving this relation and investigating other quantum geometric response properties of these systems, such as establishing a Thouless-Kohmoto-Nightingale-de Nijs (TKNN) formula \cite{PhysRevLett.49.405} for open systems is an intriguing direction for future study.

        \subsubsection{Second-Order Non-Linear Response}

\begin{figure}
                \centering
                \includegraphics[width=\linewidth]{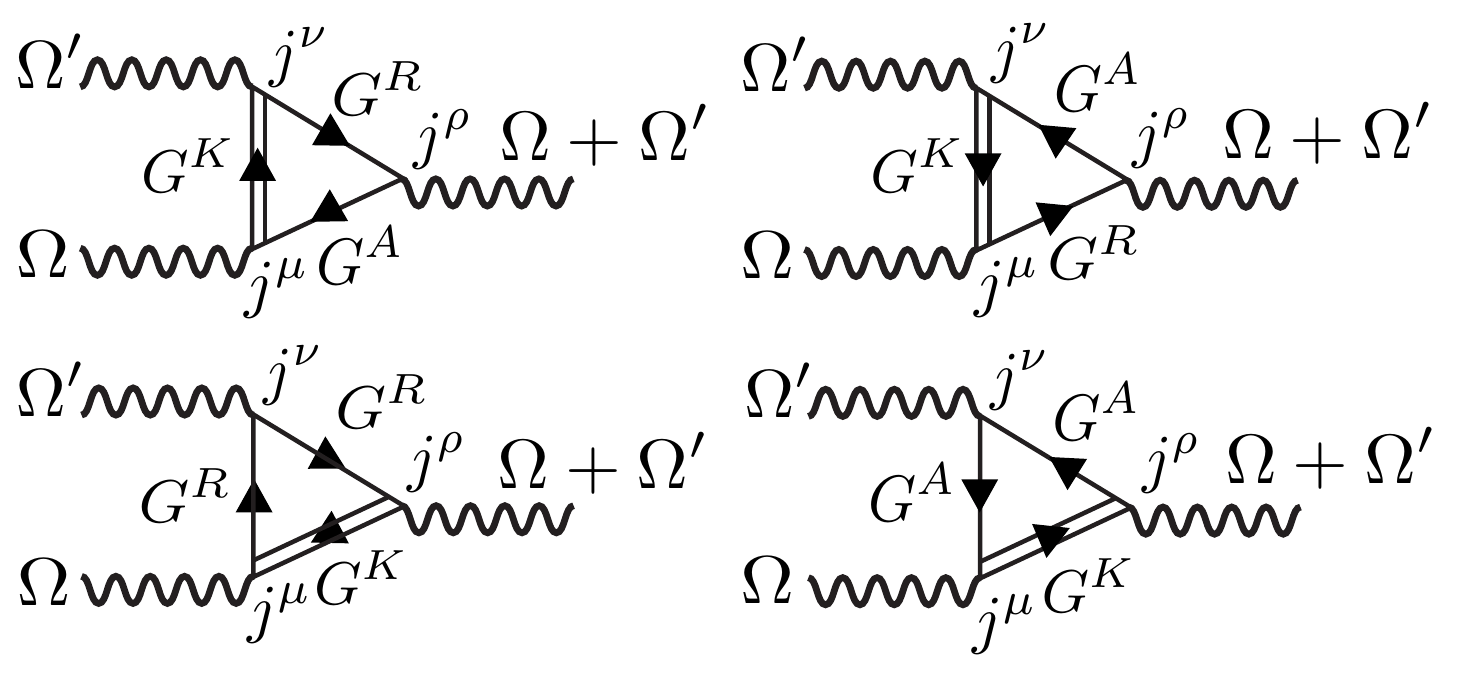}
                \caption{In second order response, a set of triangle diagrams are relevant. There are four diagrams, corresponding to the lines of Eq. (\ref{eq:triangle}), proceeding counterclockwise from the lower right. Here we illustrate the case of current response to light where $O=j^\mu$, $O'=j^\nu$ and $O''=j^\rho$.}
                \label{fig:nonlinear-feynman}
            \end{figure}

            For the second order response we have the correlation function 
            \begin{align}
                \Pi_\text{tri}(t,t') &= -i\langle [\O(0),[\O'(t),\O''(t+t')]]\rangle\theta(t)\theta(t')\nonumber\\
                &= -i[\langle \O(0)\O'(t)\O''(t+t')\rangle\nonumber\\ &- \langle \O(0)\O''(t+t')\O'(t)\rangle\nonumber\\ &- \langle \O'(t)\O''(t+t')\O(0)\rangle\nonumber\\ &+ \langle \O''(t+t')\O'(t)\O(0)\rangle]\theta(t)\theta(t'),
            \end{align}
            where the new term at this order is the triangle diagram illustrated in Fig. \ref{fig:nonlinear-feynman}.
            In analogy to the linear response, we can use Wick's theorem, express in terms of Green's functions and Fourier transform to find
            \begin{align}\label{eq:triangle}
                \Pi_\text{tri}&(\Omega,\Omega') = -i\langle [\O(\Omega),[\O'(\Omega'),\O''(\Omega+\Omega')]]\rangle\\
                =& -\frac{1}{2} e^{i\phi}\int \frac{d\omega}{2\pi}
                ~\nonumber\\ \times\bigg( &\Tr[ O_{\bm{k}} G^K_{\bm{k}}(\omega + \Omega + \Omega') O''_{\bm{k}} G^A_{\bm{k}}(\omega) O'_{\bm{k}} G^A_{\bm{k}}(\omega + \Omega') ] \notag\\
        	   +& \Tr[ O_{\bm{k}} G^R_{\bm{k}}(\omega + \Omega + \Omega') O''_{\bm{k}} G^A_{\bm{k}}(\omega) O'_{\bm{k}} G^K_{\bm{k}}(\omega + \Omega') ] \notag\\
        	   +& \Tr[ O_{\bm{k}} G^K_{\bm{k}}(\omega - \Omega') O'_{\bm{k}} G^R_{\bm{k}}(\omega) O''_{\bm{k}} G^A_{\bm{k}}(\omega - \Omega - \Omega') ]  \notag\\
        	   +& \Tr[ O_{\bm{k}} G^R_{\bm{k}}(\omega - \Omega') O'_{\bm{k}} G^R_{\bm{k}}(\omega) O''_{\bm{k}} G^K_{\bm{k}}(\omega - \Omega - \Omega') ]  \bigg),
            \end{align}
            which can be diagrammatically expressed as in Fig. \ref{fig:nonlinear-feynman}. As above, the trace includes an integral over $\bm{k}$. Now, after substituting the expression for the spectral representation and completing the $\omega$ integral we find
            \begin{widetext}
            \begin{align}\label{eq:tri_response}
            	\Pi_\text{tri}(\Omega,\Omega') =  -ie^{i\phi}\sum_{\bm{k},n_1,n_2,n_3,n_4} &\left\{ \frac{ \braOPket{u_{\bm{k},n_4}}{O_{\bm{k}}}{u_{\bm{k},n_1}}  \braOPket{\bar{u}_{\bm{k},n_1}}{\sigma^i\Sigma^K_{\bm{k}}\sigma^i}{\bar{u}_{m_2}} \braOPket{u_{\bm{k},n_2}}{O_{\bm{k}}'' \sigma^i}{\bar{u}_{\bm{k},n_3}} \braOPket{u_{\bm{k},n_3}}{O_{\bm{k}}' \sigma^i}{\bar{u}_{\bm{k},n_4}} }{ (\E_{\bm{k},n_1} - \E^*_{\bm{k},n_2}) (\Omega + \Omega' + \E^*_{\bm{k},n_3} - \E_{\bm{k},n_1}) (\Omega + \E^*_{\bm{k},n_4} - \E_{\bm{k},n_1})} \right. \notag\\
            \
            	&+ \braOPket{u_{\bm{k},n_4}}{O_{\bm{k}}}{u_{\bm{k},n_1}}  \braOPket{\bar{u}_{\bm{k},n_1}}{\sigma^i O_{\bm{k}}'' \sigma^i}{\bar{u}_{\bm{k},n_2}} \braOPket{u_{\bm{k},n_2}}{O_{\bm{k}}'}{u_{\bm{k},n_3}} \braOPket{\bar{u}_{\bm{k},n_3}}{\sigma^i\Sigma^K_{\bm{k}}\sigma^i}{\bar{u}_{\bm{k},n_4}} \times \notag\\
            		&~~~~\times \bigg( \frac{1}{(\xi_{\bm{k},n_4}^*-\xi_{\bm{k},n_3})(\Omega'+\xi_{\bm{k},n_2}^*-\xi_{\bm{k},n_3})(\Omega+\xi_{\bm{k},n_3}-\xi_{\bm{k},n_1})}\notag\\ &\qquad -  \frac{1}{(\Omega+\Omega'+\xi_{\bm{k},n_2}^*-\xi_{\bm{k},n_1})(\Omega+\xi_{\bm{k},n_3}-\xi_{\bm{k},n_1})(\Omega+\xi_{\bm{k},n_4}^*-\xi_{\bm{k},n_1})} \bigg) \notag\\
            \
            	&+ \braOPket{u_{\bm{k},n_4}}{O_{\bm{k}}}{u_{\bm{k},n_1}}  \braOPket{\bar{u}_{\bm{k},n_1}}{\sigma^i\Sigma^K_{\bm{k}}\sigma^i}{\bar{u}_{\bm{k},n_2}} \braOPket{u_{\bm{k},n_2}}{O_{\bm{k}}'}{u_{\bm{k},n_3}} \braOPket{\bar{u}_{\bm{k},n_3}}{\sigma^i O_{\bm{k}}'' \sigma^i}{\bar{u}_{\bm{k},n_4}} \notag\\
            	&~~~~\times \bigg( \frac{1}{(\xi_{\bm{k},n_2}^*-\xi_{\bm{k},n_1})(\Omega'+\xi_{\bm{k},n_1}-\xi_{\bm{k},n_3})(\Omega+\xi_{\bm{k},n_4}^*-\xi_{\bm{k},n_1})}\notag\\ &\qquad - \frac{1}{(\Omega'+\xi_{\bm{k},n_1}-\xi_{\bm{k},n_3})(\Omega'+\xi_{\bm{k},n_2}^*-\xi_{\bm{k},n_3})(\Omega+\Omega'+\xi_{\bm{k},n_4}^*-\xi_{\bm{k},n_3})} \bigg) \notag\\
            \
            	&- \left.\frac{ \braOPket{u_{\bm{k},n_4}}{O_{\bm{k}}}{u_{\bm{k},n_1}}  \braOPket{\bar{u}_{\bm{k},n_1}}{\sigma^i O_{\bm{k}}'}{u_{\bm{k},n_2}} \braOPket{\bar{u}_{\bm{k},n_2}}{\sigma^i O_{\bm{k}}''}{u_{\bm{k},n_3}} \braOPket{\bar{u}_{\bm{k},n_3}}{\sigma^i\Sigma^K_{\bm{k}}\sigma^i}{\bar{u}_{\bm{k},n_4}} }{ (\E_{\bm{k},n_3} - \E^*_{\bm{k},n_4})(\Omega + \E^*_{\bm{k},n_4} - \E_{\bm{k},n_1})(\Omega + \Omega' + \E^*_{\bm{k},n_4} - \E_{\bm{k},n_2})} \right\}
            \end{align}
            \end{widetext}
            where $\phi=\pi$ and $\sigma^i=\1$ for fermions and $\phi=2\pi$ and $\sigma^i=\sigma^3$ which acts on the Nambu structure for bosons.
            This can now be immediately evaluated in terms of the normal modes of $\sigma^i \Xi$. The derivation is given in the Supplemental Material \footnote{See Supplemental Material, Section 3 for derivations of the frequency dependent nonlinear response function.}

            \subsubsubsection{Example: Non-Linear Optical Conductivity}
            
            As an example we consider the non-linear response to light in velocity gauge, where as in the closed system the non-linear optical conductivity is given by \cite{sipe2000second}
            \begin{align}
                \sigma^{\rho\nu\mu}(\Omega,\Omega') = - \frac{1}{\Omega\Omega'}\bigg(\chi^{\rho\nu\mu}(\Omega,\Omega') + \frac{1}{2}\langle \J^{\rho\nu\mu}\rangle \bigg),
            \end{align}
            where
            \begin{align}
                \chi^{\rho\nu\mu}(\Omega,\Omega') = &-i \int_0^\infty dt\ \bigg(e^{-i\Omega t} \langle[\J^{\rho\nu}(t),\J^\mu(0)]\rangle\nonumber\\ &\qquad\!\!+ \frac{1}{2}e^{-i(\Omega+\Omega')t}\langle[\J^\rho(t),\J^{\nu\mu}(0)]\rangle\bigg)\nonumber\\
                &- \int_0^\infty dt \int_{0}^\infty dt'\ e^{-i\Omega' t}e^{-i\Omega(t+t')}\nonumber\\&\qquad\qquad \langle [\J^\mu(0),[J^\nu(t),\J^\rho(t+t')]]\rangle.
            \end{align}
            The terms including $\J^{\mu\nu\rho}$ and $\J^{\mu\nu}$ can be evaluated using the linear response formalism of the previous subsection using Eq. (\ref{eq:dia_response}) and Eq. (\ref{eq:para_response}) with $j^{\mu\nu\rho}=\partial_{k_\mu}\partial_{k_\nu}\partial_{k_\rho}H_{\bm{k}}$ and $j^{\mu\nu}=\partial_{k_\mu}\partial_{k_\nu} H_{\bm{k}}$ respectively where $\mathsf{J}$ and $j$ are related by Eq. (\ref{eq:single-particle-op}).
            The final triangle term can be evaluated using Eq. (\ref{eq:tri_response}) with $O=j^\mu$, $O'=j^\nu$, $O''=j^\rho$.

            Now, when the system is driven by one frequency of light there are two natural second order responses: second harmonic generation/frequency doubling, and shift current generation/frequency cancellation that are given by \cite{sipe2000second}
            \begin{align}
            \sigma^{\mu\nu}_\text{2HG}(\Omega) = \sigma^{\mu\mu\nu}(\Omega,\Omega),
            \end{align}
            and 
            \begin{align}
            \sigma^{\mu\nu}_\text{shift}(\Omega) = \tfrac{1}{2}[\sigma^{\mu\mu\nu}(\Omega,-\Omega)+\sigma^{\mu\mu\nu}(-\Omega,\Omega)].
            \end{align}
            which can be naturally evaluated using the formulae above.

    \subsection{Applications} \label{sec:examples}

        Having established a response formalism for quadratic Lindbladians, we now apply it to diverse physical examples illustrating its applicability. In the first example we study an XY spin chain which through the Jordan-Wigner transformation can be expressed in terms of free fermions. This fermionic model has superconducting terms and leverages the full Nambu structure of the theory. In the second example, we consider the paradigmatic material Bernal bilayer graphene and its linear and non-linear optical responses. In the third example, we consider a bosonic optical lattice and consider the momentum space atomic occupation that results from realistic slight anisotropies in dissipation rates.

        \subsubsection{XY Spin Model}

            In equilibrium, the 1D $J_x, J_y$ (XY) spin-1/2 model exhibits three phases: an oscillatory spin-density wave (SDW) phase at small fields and small $J^x/J^y$ anisotropy, an ordered ferromagnetic (FM) phase at small fields and large anisotropy, and a disordered paramagnetic (PM) phase at large fields \cite{franchini2017introduction}. Recently the Ising limit of this model has been realized on a quantum computer where simulated dissipation cooled the system towards its ground state \cite{mi2023stable}. Remarkably, in the dissipatively boundary-driven XY model similar phases to equilibrium can emerge \cite{prosen2008quantum,eisert2010noise,prosen2010exact,landi2022nonequilibrium}, however their signatures in physical spectroscopic responses and consequences on magnetic excitations are an open question. To address this, we recast the boundary driven model as a quadratic Lindbladian and compute the dynamical spin response at finite transverse field strengths. We find that, in contrast to the gapped modes of the equilibrium model, there are gapless modes whose coupling to the single site spin flip operator $S_i^z$ decreases as the spin-density wavelength increases.

            \begin{figure*}
                    \centering
                    \includegraphics[width=\linewidth]{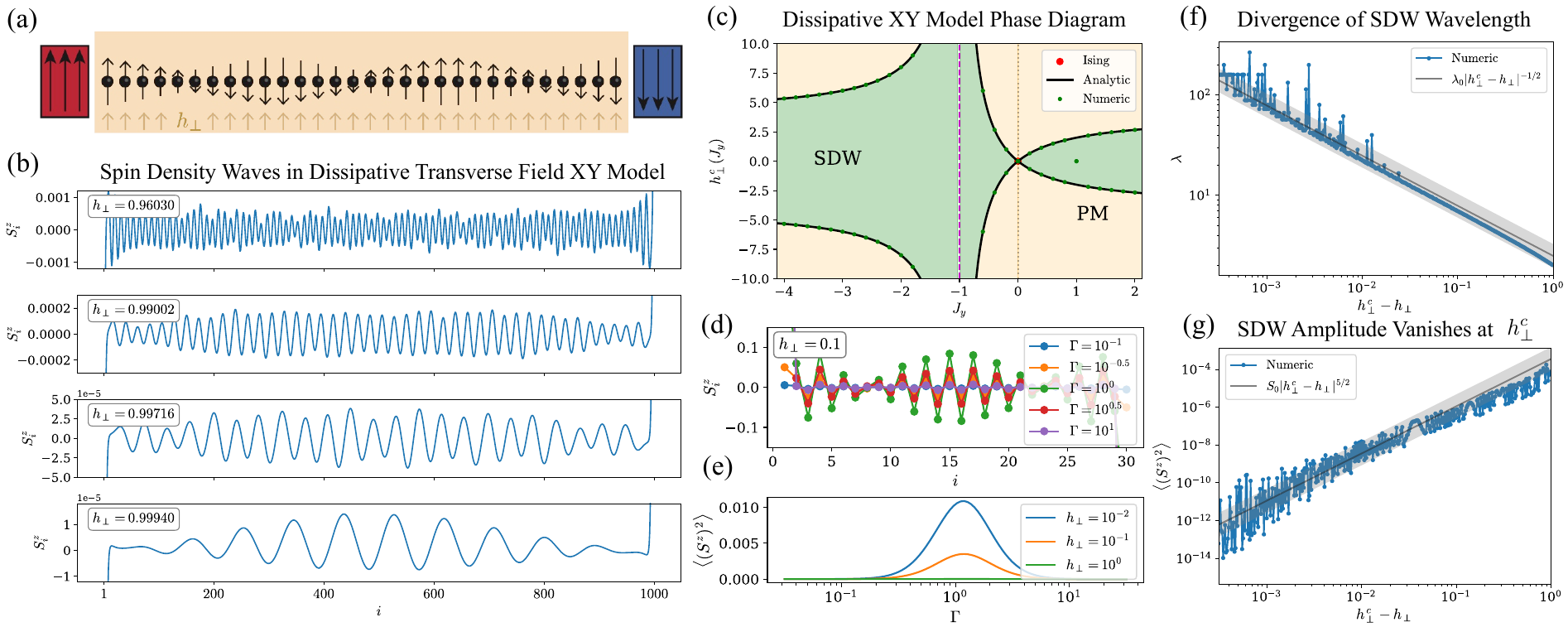}
                    \caption{The transverse-field XY model with boundary dissipation exhibits spin density wave (SDW) and paramagnetic (PM) phases where the wavelength of the spin density waves diverges at the critical point; the SDW amplitude depends on the dissipation rate and vanishes at the critical point. \textbf{(a)} Schematic of an XY chain coupled to ferromagnetic reservoirs on the boundaries. \textbf{(b)} SDWs near criticality in the XY model with $J_x=1$ and $J_y=1/3$ for $\Gamma=1$ and $N=1000$. \textbf{(c)} Phase diagram; at large fields the phase is paramagnetic except at $J_y=-J_x$  (purple line) where there is only a SDW phase because of a ``seesaw" mechanism. At the Ising point, $J_y=0$ (dotted line), the oscillatory phase vanishes. Additionally, precisely at $J_x=J_y$ the system is the XX critical model which does not exhibit an oscillatory phase; here, $J_y$ is measured in units of $J_x$, i.e. $J_x=1$. \textbf{(d-e)} SDW amplitude depends on the dissipation rate and peaks when $\Gamma\sim J$ so that transitions from the reservoir to the second site is maximized as understood using second order perturbation theory, illustrated for a $N=30$ chain with $J_x=1$, $J_y=1/3$ \textbf{(f)} SDW wavelength diverges at the critical point with critical exponent $-1/2$, illustrated for $N=1000$, $J_x=\Gamma=1$, $J_y=1/3$ although the exponent is the same for different parameter choices. \textbf{(g)} The SDW amplitude vanishes at the critical point, independent of $\Gamma$, and this dominates the spin-spin correlations.}
                    \label{fig:spin}
                \end{figure*}

            \subsubsubsection{Model}
    
                Consider first the paradigmatic $J_x, J_y$ spin chain model with $N$ sites and open boundary conditions
                \begin{align}
                    H_{XY} = \sum_{n=1}^{N-1} J^x S_n^x S_{n+1}^x + J^y S_n^y S_{n+1}^y .
                \end{align}
                Additionally we consider the effect of local potentials (transverse fields) $h_n$
                \begin{align}
                    H_\perp = \sum_{n=1}^{N} h\, S_n^z,
                \end{align}
                so that we have $H=H_{XY}+H_\perp$.
                The standard Jordan-Wigner transformation recasts this model in terms of free fermions
                \begin{align}
                    S_n^x &= \tfrac{1}{2}(S_n^++S_n^-)\\ 
                    S_n^y &= \tfrac{1}{2i}(S_n^+-S_n^-)\\
                    S_n^+ &= e^{-i\pi \sum_{m<n}c_m^\dagger c_m} c_n^\dagger\\
                    S_n^- &= e^{i\pi \sum_{m<n}c_m^\dagger c_m} c_n\\
                    S_n^z &= c_n^\dagger c_n - \tfrac{1}{2},
                \end{align}
                with a single-particle Hamiltonian where we multiplied the $J_x$ and $J_y$ terms by 2 so that the critical transverse field in the Ising limit of the closed system is at $h_\perp = 1$
                \begin{align}
                    H_{XY}
                    &= \frac{1}{2} \sum_{n=1}^{N-1} (J^x-J^y) c_n^\dagger c_{n+1}^\dagger + (J^x + J^y) c_n^\dagger c_{n+1} + h.c. \\
                    H_\perp &= \sum_{n=1}^{N} h (c^\dagger_nc_n-\tfrac{1}{2}).
                \end{align}

                Now, suppose that the model is subjected to boundary dissipators given by the jump operators
                \begin{align}
                    J_1 &= \Gamma_1 c^\dagger_1 = \Gamma_1 S^+_1\\
                    J_N &= \Gamma_N c_N = \Gamma_N S^-_N,  
                \end{align}
                where the Jordan-Wigner string for $J_N$ on the two contours cancels as the total parity of the superfermions is conserved---annihilation of one fermion on one contour corresponds to annihilation of one fermion on the other contour and so the total partiy is preserved. The sign is unimportant for the Lindblad evolution since each jump operator appears with its adjoint. These jump operators describe the coupling of the ends of the XY chain to two oppositely spin-polarized leads. These will preferentially (and incoherently) attempt to collapse the spins at the two ends of the chain to $\uparrow$ and $\downarrow$ respectively.
                The model is depicted schematically in Fig. \ref{fig:spin}(a) where a 1D chain is subjected to a transverse field and is sandwiched in between ferromagnetic reservoirs.

            \subsubsubsection{Magnetization Density}

                We are now in a position to calculate the steady-state $z$-polarization magnetization density, $S^z=n-\tfrac{1}{2}$, by using the fermionic response formalism given in Eq. (\ref{eq:density}) and above.  We find a steady-state SDW phase with mean magnetization zero and long-range spin density wave order as illustrated in Fig. \ref{fig:spin}(b), and a paramagnetic (PM) phase with mean magnetization zero and no long-range order: spin-spin correlations decay exponentially.                
                The critical line between the SDW and the PM phases of the dissipative chain is at
                \begin{align}
                    h_\perp^c(J_y) = \frac{4 J_xJ_y}{J_x+J_y}.
                \end{align}
                
                Notably, the location of the critical line is independent of the dissipation rate $\Gamma$. While as illustrated in Fig. \ref{fig:spin}, $\Gamma$ plays a key role in determining the amplitude of oscillations in the SDW phase, the oscillations only \textit{exactly} vanish at and above the critical field value in the PM phase, independent of $\Gamma$.
                This critical field can be calculated from inspecting the \textit{bulk} Jordan-Wigner spectrum of the XY spin chain Hamiltonian
                \begin{align}
                \epsilon(k) = 2(J_x\!+\!J_y) \sqrt{\left(\!\frac{h}{J_x\!+\!J_y}\!-\!\cos(k)\!\right)^2 \!\!\!\!+\! \bigg(\!\frac{J_x\!-\!J_y}{J_x\!+\!J_y}\!\bigg)^2\!\!\!\sin^2(k)}
                \end{align}
                For large transverse fields $h$ in the PM phase, the Jordan-Wigner dispersion has a minimum at $k=0$. Conversely, for small transverse fields $h$ in the SDW phase, the dispersion exhibits two minimal at $k = \pm k_{\rm SDW}$ with
                \begin{align}
                    k_{\rm SDW} = \arccos\left( \frac{h(J_x+J_y)}{4 J_x J_y} \right)
                \end{align}
                The jump between $k=0$ and $k=\pm k_{\rm SDW}$ dispersion minima occurs at $h_\perp^c(J_y) = \frac{4 J_xJ_y}{J_x+J_y}$ and signifies the phase transition upon adding the boundary dissipation. As the phase boundary is crossed from PM to SDW, the dissipative steady state can now preferentially populate the $\pm k_{\rm SDW}$ bulk modes, resulting in a steady state SDW order with momentum $k_{\rm SDW}$. We note that this mechanism relies on an intrinsically non-equilibrium steady-state distribution, and stands in contrast to the ferromagnetic to paramagnetic transition in closed equilibrium systems which occurs when the spectrum becomes gapless, i.e. $\epsilon(0)=0$ at $h_\perp^c(J_y)=J_y+J_x$ \cite{franchini2017introduction}.

                We illustrate the phase diagram in Fig. \ref{fig:spin}(c) and note key features and symmetries of the phase diagram. First, at the Ising point $J_x=1$, $J_y=0$ there is no spin-density wave phase as there is only a non-magnetized phase. Second, at the finely tuned point $J_y=J_x$, the model is a critical XX spin chain and the SDW phase vanishes. Third, at $J_y=-J_x$, the SDW phase persists for all transverse fields: this can be understood through a seesaw mechanism where ferromagnetic correlations in one direction are precisely counterbalanced by correlations in the opposite direction leading to persistent oscillatory behavior, even at strong fields. Moreover the critical line of the model exhibits two symmetries: first, there is $\mathbb{Z}_2$ symmetry corresponding to spin-flip symmetry, and second there is a glide symmetry: upon reflecting the phase boundary lines across the seesaw point (purple line) in Fig. \ref{fig:spin}(c) they are the same as those on the other side of the seesaw point before reflection up to a shift in $h_\perp$ by $\pm 8J_x$.
    
                While the dissipation rate $\Gamma$ does not affect the wavelength of the SDW oscillations, it has a drastic effect on the excitation spectrum, changing the amplitude of the oscillations as we illustrate in Fig. \ref{fig:spin}(d-e). The amplitude of the SDW oscillations is maximized when $\Gamma\sim J$ where $J$ is the nearest neighbor spin-spin coupling. This can be understood using second order perturbation theory, where we consider the first spin in the system. This spin is coupled both to a ferromagnetic reservoir by $\Gamma$ and to the rest of the system (more or less paramagnetic) by $J$. When $\Gamma<\!\!<J$, the coupling to the center of the system dominates and oscillations are weak. When $\Gamma>\!\!>J$ there are no virtual transitions (via second order perturbation theory) from the first spin to the second spin and so oscillations are weak. Finally, when $\Gamma\sim J$, the amplitude of oscillations peaks since there is both a strong drive and a strong coupling to the next spin in the chain which means that the virtual transitions from perturbation theory are significant.

                \begin{figure*}
                    \centering
                    \includegraphics[width=\linewidth]{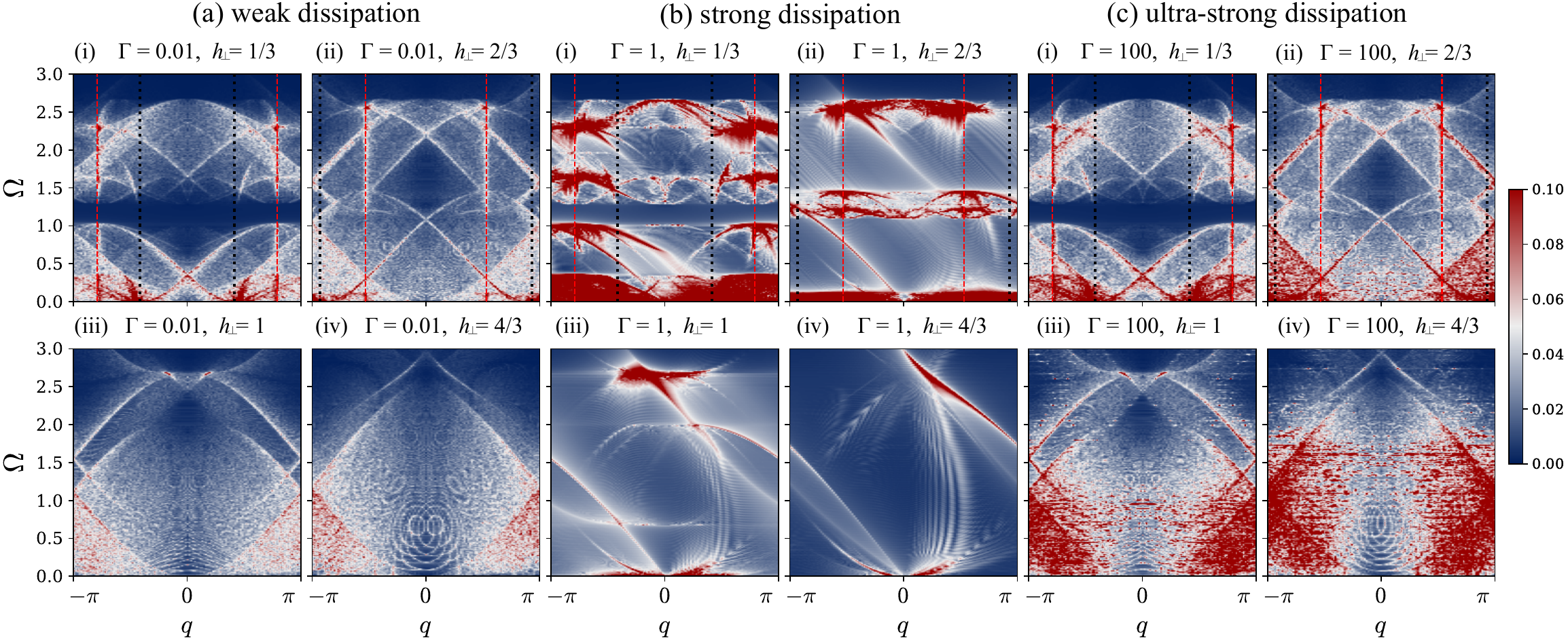}
                    \caption{Dynamic spin susceptibility, $|\mathcal{S}(q,\Omega)|$ of the boundary driven XY model with $J_x=1$ and $J_y=1/3$ with $N=301$ sites. The critical point for spin-density wave (SDW) order is at $h_\perp=1$ independent of $\Gamma$. Sub-panels (i) and (ii) correspond to small transverse fields where SDW wave order exists. Sub-panels (iii) are at the critical point, and Sub-panels (iv) are at strong fields in the paramagnetic regime. Additionally, we compare $\Gamma=0.01$, $1$ and $100$ in panels (a), (b), and (c) respectively. In panels (b), $\Gamma=1\sim J$ maximizes the magnitude of the spin-spin correlation and for the $\Gamma=1$ panels we divide $\mathcal{S}$ by 10 so that we can use a uniform color scale across all panels. In this case inversion symmetry is broken by the bath as effects of the boundary dissipation are felt deep in the bulk of the system. Choosing $\Gamma$ much smaller (a) (or larger (c)) than $J$ leads to an approximate restoration of inversion symmetry as the effects of dissipation are primarily localized to the edges of the system. This can be understood in terms of Fig. 4(e) where second-order perturbation theory dictates that the amplitude of tunneling from the first site to the second site (and further into the bulk) is maximized for $\Gamma\sim J$. In contrast to the \textit{gapped} excitations of the closed system, the dissipative system hosts \textit{gapless} dispersive excitations as the Lindbladian steady state exhibits long-wavelength SDW fluctuations on top of a paramagnetic background. Wavevectors $\pm q_\mathrm{SDW}$ are illustrated with dashed red dashed vertical lines. Wavevectors $\pm2q_\mathrm{SDW} ~\textrm{mod}~ 2\pi$ are illustrated with black dotted vertical lines correspond to gapless modes dispersing from zero frequency.}
                    \label{fig:spin-qw}
                \end{figure*}

                Near the critical point, the \textit{wavelength} of the spin density wave diverges. We attribute a critical exponent to this divergence and find $-1/2$ as expected for a free theory. This critical exponent matches the exponent for correlation functions defined by Eisert and Prosen in Ref. \cite{eisert2010noise}. We illustrate this power law divergence in Fig. \ref{fig:spin}(f), with noise for long wavelengths arising when commensurability with the system size ($N=1000$) becomes significant. Far from $h_\perp^c$, there are deviations from the power law behavior. The wavelengths $\lambda$ are extracted by Fourier transforming the $z$-magnetization density of the central region of the spin chain and selecting the frequency of the largest peak.
                
                Now, it is significant to note that while the wavelength of the SDW diverges at the critical point, its amplitude vanishes as we illustrate in Fig. \ref{fig:spin}(g). Near the critical point for any $l<\!\!<\lambda$ for spin-density wavelength $\lambda$ we have
                \begin{align}
                \langle (S^z)^2\rangle
                = \frac{1}{L}\sum_n (S_n^z)^2 &= \lim_{h_\perp\to h_\perp^c} \langle |S^z_n S^z_{n+l}|\rangle
                \end{align} 
                since the correlation function is over spins near each other on one wavelength and hence have asymptotically equal magnetizations. The average is over all sites $n$, and for numeric simulations we take $l=2$.
                Studying this numerically, we find the amplitude vanishes with a critical exponent of roughly $5/2$. The data is noisy over the whole range studied because the amplitude oscillates with the constructive and destructive interference of SDWs originating from the left and right ferromagnetic reservoirs. Again, far from the critical point we see deviations from power-law behavior.

            \subsubsubsection{Dynamic Spin Susceptibility}

                Above we considered the steady state magnetization, now we consider the dynamic response of the spin chain to a frequency dependent perturbation. This is analogous to calculations of time-dependent spin correlation functions in \cite{kos2017time,wang2022exact}, except that in our framework frequency rather than time is the natural footing and so there is no need to Fourier transform over a finite time to obtain the dynamic spin susceptibility. Here we specialize to the case of an XY model with $J_x=1$, $J_y=1/3$ that exhibits both SDW and paramagnetic phases. As in the static case, the amplitudes peak when $\Gamma\sim J$, so we choose $\Gamma=1$.
                
                We proceed to calculate the dynamic spin-spin susceptibility using Eq. (\ref{eq:paramagnetic}) for the one-loop finite frequency response. We consider the operators $O=S_i^z$ and $O'=S_j^z$ where $i$ and $j$ are site indices in the chain. We identify $\mathcal{S}(i,j,\Omega)=\Pi^{ij}(\Omega)$ as the spin-spin correlation function, and compute the specific correlations $\mathcal{S}(2i,\Omega)=\mathcal{S}(N/2-i,N/2+i,\Omega)$ which is the spin-spin correlation function centered around the middle of the chain, where $i$ ranges from 0 to $N/2-\epsilon$ where we choose $N=301$ and $\epsilon=25$ to avoid finite size boundary effects. We then Fourier transform this to obtain $\mathcal{S}(q,\Omega)$. Note that
                $\langle [S_j^z,S_i^z]\rangle
                = \langle [n_j,n_i]\rangle
                $ so we are free to complete this computation using fermions.

                Visualizing $\mathcal{S}(q,\Omega)$ in Fig. \ref{fig:spin-qw}, we see that there are gapless modes in contrast to the closed system where all modes are gapped for $h_\perp\leq h_\perp^c$. Notably, while the real part of the spectrum of $\Xi$ would suggest that the open system has a gap above the $\mathrm{Re}(\xi)=0$ modes there are gapless excitations because the modes are partially filled as given by the distribution function. In fact, for our choice of jump operators \textit{all} states are half-filled except four states corresponding to boundary modes acted on directly by the dissipators. In Fig. \ref{fig:spin-qw}(a)(i) and Fig. \ref{fig:spin-qw}(c)(i) a gap in the spectrum of $\mathcal{S}(q,\Omega)$ appears between $\Omega=1$ and 1.2, corresponding to the gap in the spectrum of $\Xi$: while transitions are still allowed within the lower (and upper) manifold of states, the gap from the lower to the upper manifold exceeds the bandwidth of the lower (and upper) manifold and so no transitions are possible at these frequencies. In contrast in Fig. \ref{fig:spin-qw}(a)(ii) and Fig. \ref{fig:spin-qw}(c)(ii) no gap is visible since the bandwidth of the manifolds exceeds the gap between them.
                
                In the weak and ultra-strong dissipation limits,  Fig. \ref{fig:spin-qw}(a) and Fig. \ref{fig:spin-qw}(c) respectively, gapless modes disperse from $\pm 2q_\mathrm{SDW}~\mathrm{mod}~2\pi$ for spin-density wavenumber $q_\mathrm{SDW}$ which exists for $h_\perp\leq h_\perp^c = 1$ here. Above $h_\perp^c$ closed systems similarly exhibit a gapless response. Additionally for $h_\perp\leq h_\perp^c$ we note that there are excitations at $q_\mathrm{SDW}$ corresponding to exciting spin-density waves.
                The weak dissipation and ultra-strong dissipation cases have comparable spectra with a clear correspondence between the dominant modes because, as illustrated in Fig. \ref{fig:spin}(e), for $\Gamma$ not comparable to $J$ coupling to the bulk is weak and the effect of dissipation is felt most strongly near the boundaries.

                In the strong dissipation case, at commensurate $\Gamma\sim J$, inversion symmetry instead becomes strongly broken as the effects of dissipation persist deep in the bulk. At $h_\perp=J_x/3$ and $h_\perp=2J_x/3$, as visualized in Fig. \ref{fig:spin-qw}(b)(i-ii) the single spin flip operators $O=S_i^z$ and $O'=S_j^z$ couple strongly to a series of dispersive modes that are gapless. Meanwhile at larger field strengths, as illustrated in Fig. \ref{fig:spin-qw}(b)(iii-iv), the operators couple more weakly to the gapless dispersive modes. This is also the regime in which the spin-density wave order of the steady state has a long wavelength. This may be because the ideal description of the excitations is as a more-extended object and the single-spin flip probe is too local to efficiently probe the excitations. Above $h_\perp^c$, Fig. \ref{fig:spin-qw}(b)(iv), we see that at large energies there is only one mode with significant spectral weight, while in Fig. \ref{fig:spin-qw}(b)(ii) there are many modes. We interpret this as modes that are scattered by multiples of $q_\mathrm{SDW}$ (and backfolded by $2\pi$) in the spin-density wave phase, but that collapse onto a single mode above $h_\perp^c$ when the spin-density wave order vanishes and $q_\mathrm{SDW}$ goes to 0.

        \subsubsection{Bilayer Graphene}

\begin{figure*}
                \centering
                \includegraphics[width=\linewidth]{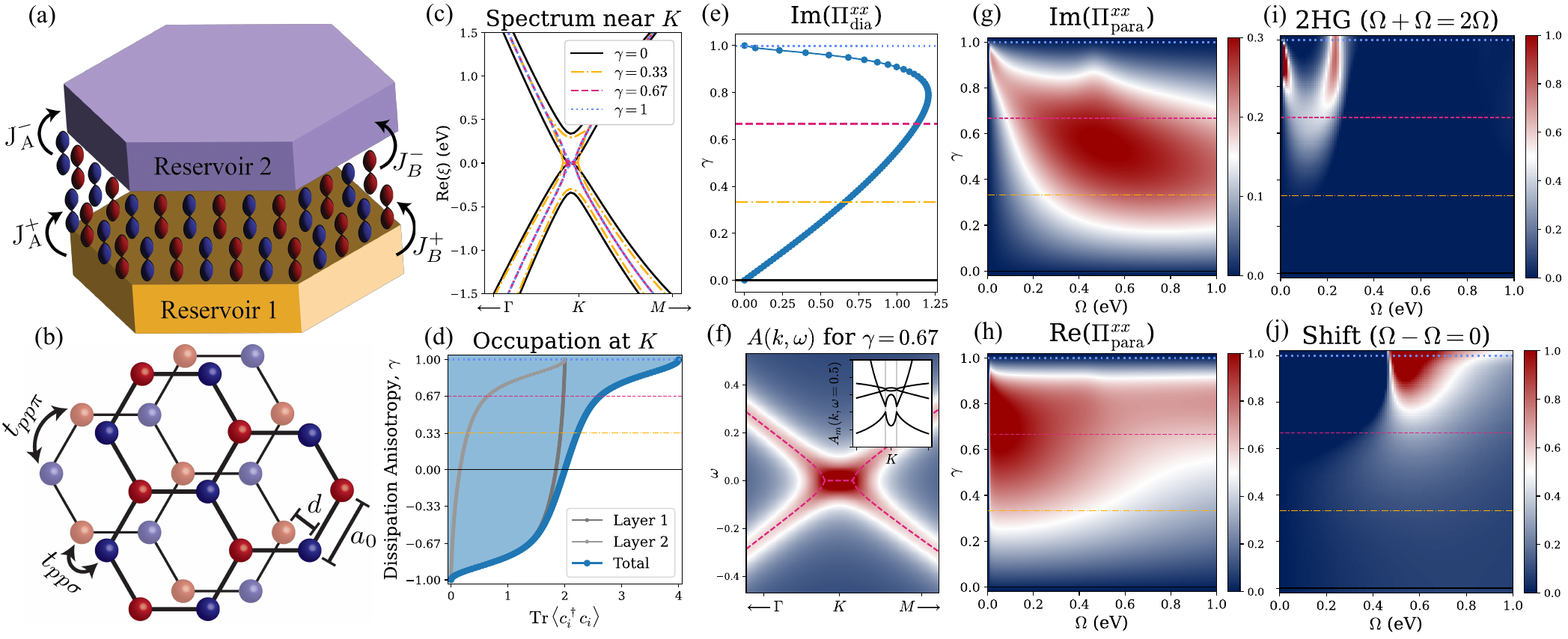}
                \caption{Structure, occupation, and optics of Bernal bilayer graphene dissipatively coupled two reservoirs. Note that the linear optical conductivity is $\sigma=i\Pi/\Omega$ and the non-linear optical conductivity is $\sigma=-\Pi/\Omega^2$ and all conductivities are measured in units of $e^2/\hbar$. \textbf{(a)} Schematic illustration of the system and system-reservoir coupling $J$. \textbf{(b)} Top view of the bilayer with tight-binding parameters $t$ and distances $d$ and $a_0$. \textbf{(c)} Spectrum of $\Xi$ near the $K$ point for select values of dissipation anisotropy $\gamma$ illustrated with black, yellow, red, and blue lines which are used in the subsequent panels. \textbf{(d)} Near $\gamma=-1$ only layer 1 (bottom) is filled while near $\gamma=1$ both layers fill. \textbf{(e)} Diamagnetic current-current correlation function, notice the decrease near $\gamma=1$ corresponds to a fundamentally open-system behavior since occupation increases monotonically with $\gamma$ and in closed systems the diamagnetic response is the density. \textbf{(f)} Momentum and energy resolved spectral function for $\gamma=2/3$; the inset is the state resolved spectral function which resolves the presence of exceptional points. \textbf{(g-h)} Paramagnetic linear optical conductivity; the onset frequency is roughly the gap at $K$ and the turn-off frequency is when high-energy bands broaden and have very short lifetimes. \textbf{(i)} Second harmonic generation, $-\text{Re}(\Pi^{xxx}_\text{2HG})$, appears when centrosymmetry is strongly broken by the coupling to the reservoirs and exhibits two peaks corresponding to transitions between pairs of bands that are close together (near $M$) and far apart (near $K$). \textbf{(j)} Shift generation, $-\text{Re}(\Pi^{xxx}_\text{shift})$, also appears at large $\gamma$ and the peak corresponds to a cycling $\Omega-\Omega=0$ process near the $K$ point.}
                \label{fig:bernal-bilayer}
            \end{figure*}

            Over the past decade, twisted bilayer graphene has emerged as a platform of choice for realizing many unusual physical phenomena such as superconductivity \cite{cao2018unconventional,yankowitz2019tuning}, and many other correlated electronic behaviors \cite{andrei2020graphene} in flat bands near the Fermi energy \cite{trambly2010localization,bistritzer_moire_2011}. The simplest and most thermodynamically stable graphene bilayer is AB stacked ``Bernal" bilayer. Recently, superconductivity \cite{zhou2022isospin} and the quantum anomalous Hall effect \cite{geisenhof2021quantum} have been observed in Bernal bilayer graphene. Here we take a new approach by strongly coupling bilayer graphene to metallic top and bottom leads and instead analyze its dissipative steady states due to potential differences between the leads. Importantly, dissipation will alter the single-particle electronic properties and lead to the formation of exceptional points. Given this fundamentally non-Hermitian phenomenon, an interesting question is how these properties are manifested in linear and non-linear optical response properties that can be probed in experiments. We illustrate the system geometry in Fig. \ref{fig:bernal-bilayer}(a) where the bottom layer is coupled to reservoir 1 and the top layer is coupled to reservoir 2.

            \subsubsubsection{Model}

                We start from a tight-binding model for the graphene bilayer; we consider the Bernal structure as illustrated in Fig. \ref{fig:bernal-bilayer}(b). The hopping terms are $t(\bm{r}) = t_{pp\pi}(\bm{r}) + t_{pp\sigma}(\bm{r})$ where the terms are given by \cite{trambly2010localization,moon2013optical}
                \begin{align}
                    t_{pp\pi}(\bm{r}) &= V_{pp\pi} e^{-(|\bm{r}|-a_0)/\delta} \left(1-\left(\frac{\bm{r}\cdot \bm{e}_z}{|\bm{r}|}\right)^2\right)\\
                    t_{pp\sigma}(\bm{r}) &= V_{pp\sigma} e^{-(|\bm{r}|-d)/\delta}\left(\frac{\bm{r}\cdot \bm{e}_z}{|\bm{r}|}\right)^2,
                \end{align}
                which are the terms that emerge using the Slater-Koster method for the $p_z$ orbitals. The energies that set the problem scale are $V_{pp\pi} = 2.7$ eV and $V_{pp\sigma} = -0.48$ eV, and the key lengths are the intra-layer spacing $a_0 = 0.142$ nm, inter-layer spacing $d = 0.335$ nm, and exponential decay length $\delta = 0.0453$ nm, finally $\bm{e}_z = (0,0,1)$ is the unit vector perpendicular to the bilayer. We implement a cutoff for the tight-binding hopping so that $t(\bm{r})=0$ when $|\bm{r}|>1$ nm.

                Next, we introduce spatially (sublattice) homogeneous, but layer dependent dissipation given by
                \begin{align}
                J^+_{A/B} = \Gamma(1+\gamma) c_{1A/B}^\dagger\\
                J^-_{A/B} = \Gamma(1-\gamma) c_{2A/B},
                \end{align}
                where the first layer is subjected to particle gain with amplitude $\Gamma(1+\gamma)$ and the second layer is subjected to particle loss with amplitude $\Gamma(1-\gamma)$ as illustrated in Fig. \ref{fig:bernal-bilayer}(a). $\Gamma$ sets the overall scale of dissipation, and $\gamma$ sets an anisotropy in coupling between the lower layer to the lower reservoir and the upper layer to the upper reservoir. This corresponds to a bilayer sandwiched between two top and bottom reservoirs at different chemical potentials, where we neglect electric field effects induced by charge distributions in the reservoirs.

                To emphasize the effect of dissipation, we will set $\Gamma$ to 1 eV, to compete with energy scales of the (coherent) closed-system band structure; while perhaps exceedingly large for Bernal bilayers, similar physics occurs in large twist angle commensurate bilayer graphene but at a greatly reduced energy scales \cite{shallcross2008quantum,mele2010commensuration,talkington2023electric,talkington2023terahertz}. This reduced energy scale means that in such large twist angle samples the relevant low energy system physics may occur at the same energy as a reduced system-reservoir coupling $\Gamma.$  

                Now, we observe some interesting features of the dissipation acting on the bilayer as exhibited in Fig. \ref{fig:bernal-bilayer}(c). First, for zero anisotropy $\gamma$, dissipation acts solely to induce an overall broadening for electronic states. As a finite anistropy $\gamma$ between tunneling into the high-bias and low-bias dissipative leads is induced, a non-equilibrium steady state forms, in which particles tunnel from top to bottom gate via the Bernal stack. Here, exceptional points form near the band maxima (where the bands are almost degenerate), and then migrate towards the $K$ point as the anisotropy is increased. We can understand this by considering the bands away from the touching to be living in each layer and then hybridizing as they approach the crossing at $K$.
                Fig. \ref{fig:bernal-bilayer}(d) shows the steady state band occupation, which varies from completely empty at $\gamma=-1$ corresponding to all loss processes (i.e., the high-bias gate is cut off from the sample) to completely filled at $\gamma=1$ corresponding to all gain processes. In between these we see that the layer resolved occupation is quite anisotropic, where even at $\gamma=0$ almost all the occupation is in layer 1 where the tunneling-in processes occur and layer 2 only becomes significantly occupied near $\gamma=1$. We note that the difference in layer occupation corresponds to the rate $V_{pp\sigma}$ at which the layers are coupled where at large $V_{pp\sigma}$, the layers have the same occupation.

            \subsubsubsection{Absence of Exceptional Points in the Spectral Function}

                A defining characteristic of non-Hermitian systems is possibility for the formation of exceptional points where eigenvalues and vectors coalesce and no longer form a complete basis. While extensively discussed in \textit{classical} systems such as photonic crystals \cite{bergholtz2021exceptional}, signatures of exceptional points in open \textit{quantum} systems are less clear. 
                
                To shed light on their spectroscopic signatures, we start from discussing their impact on the single-particle spectral function
                \begin{align}
                    A_{\bm{k},n}(\omega) = -\frac{1}{\pi}\text{Im}\bigg(\text{Tr} \frac{|u_{\bm{k},n}\rangle\langle\bar{u}_{\bm{k},n}|}{\omega-\xi_{\bm{k},n}}\bigg).
                \end{align}
                While the spectral function exhibits a non-analyticity in $\omega$ at the exceptional point in each eigenstate $n$ that coalesces, the full spectral function $A=\sum_m A_m$ remains a smooth function of $\omega$. We illustrate the eigenstate-resolved non-analyticity in the inset of Fig. \ref{fig:bernal-bilayer}(f), meanwhile in the main panel of Fig. \ref{fig:bernal-bilayer}(f) we plot the full spectral function. The reason this non-analyticity is not inherited by the full spectral function
                is a fine cancellation due to contour-reversal symmetry where while differences of $\xi$ terms lead to non-analyticities, these non-analyticities \textit{cancel} when $n$ is summed over. This property of exceptional points showing up in state-resolved responses but vanishing when all states are summed over is inherited by the other response properties as well.

                \subsubsubsection{Diamagnetic Optical Conductivity}

                We now turn to the diamagnetic optical response, which takes the form
                \begin{align}
                    \sigma^{\mu\nu}_\text{dia}(\Omega) = -\frac{1}{\Omega} \sum_{\bm{k},m,m'} \frac{\langle\bar{u}_{\bm{k},m}|\Sigma^K_{\bm{k}}|\bar{u}_{\bm{k},m'}\rangle}{\xi_{\bm{k},m'}^* - \xi_{\bm{k},m}} \langle u_{\bm{k},m'}|j^{\mu\nu}_{\bm{k}}|u_{\bm{k},m}\rangle,
                \end{align}
                where for the Bernal bilayer the current operator is traceless so we drop the trace term and $j^{\mu\nu}_{\bm{k}}=\partial_{k_\mu}\partial_{k_\nu}H_{\bm{k}}$, and $\phi=\pi$ and $\sigma^i=\1$ for fermions.

                Notably as illustrated in Fig. \ref{fig:bernal-bilayer}(e), the diamagnetic optical conductivity decreases above $\gamma\sim 0.79$  which corresponds to a decreasing diamagnetic optical conductivity with increasing filling. This is notable since in closed systems the diamagnetic optical conductivity always increases monotonically with filling. So the decrease at large $\gamma=1$ is an indication of the open nature of this system.

                We note that in this case the optical conductivity is purely real (as in closed systems) which corresponds to the fact that the current operators here are Hermitian. Hence since $\Sigma^K$ is Hermitian, Eq. (\ref{eq:para_response}) is equal to its conjugate. This is not the case when the current operators are non-Hermitian and then the response is both reactive and dissipative as opposed to just being reactive as in closed systems.

            \subsubsubsection{Paramagnetic Linear Optical Conductivity}

                We now study the paramagnetic contribution to optical conductivity, which again has both a real and an imaginary contribution in dissipative quantum systems. In Fig. \ref{fig:bernal-bilayer}(h) we see that the current-current correlation is relatively constant over a range of frequencies and anisotropies. Around this region of constant response there is an onset at a frequency that seems to correspond roughly to the energy separation to the remote bands at the $K$ point, and this onset goes to zero frequency as $\gamma\to 1$. On the other hand, the response vanishes above a threshold corresponding to high energy bands being extremely short lived (large imaginary part/broadening). Finally, we notice that around $\gamma=1$ and $\Omega=0.5$ eV there is a slight crest in the optical conductivity which we attribute to some transitions near the $K$ point.
        
            \subsubsubsection{Non-Linear Optical Conductivity}

                Now, we can use our non-linear response formalism to calculate the second harmonic and shift responses. For clarity we only include the contributions from the triangle diagram and not from the other paramagnetic and diamagnetic loops. Since the closed system is centrosymmetric, the second order response vanishes, as seen in Fig. \ref{fig:bernal-bilayer}(i-j) at $\gamma=0$. For clarity, we have limited the non-linear response calculation to the triangle diagram term from Eq. (\ref{eq:triangle}), since the other contributions are qualitatively similar to the linear response. As $\gamma$ increases, the dissipative couplings to the top and bottom reservoirs break inversion symmetry, leading to a non-centrosymmetric dissipative steady state. Consequently, the non-linear response increases but doesn't peak until $\gamma\sim 0.8-1.0$ where the layers are both nearly filled.
                
                In the second harmonic response, Fig. \ref{fig:bernal-bilayer}(i), we see two peaks, one corresponding to ``narrow" interband transitions between neighboring pairs of bands around the $\Gamma$ and $M$ points, and one corresponding to ``wide" interband transitions around $K$. To understand this response it is important to note that the distribution function of the non-equilibrium steady state is \textit{not} that of a Fermi-Dirac distribution function but rather each of the bands is partially filled to some non-negligible extent.

                Meanwhile in the shift response, Fig. \ref{fig:bernal-bilayer}(j) we see the complementary behavior where there is no response until an energy threshold corresponding to cycling at the $K$ point. The onset of the shift response is rapid and it reaches a large magnitude, perhaps suggesting that using dissipation to break centrosymmetry is a promising direction for realizing devices with non-linear optical responses such as the shift current that enables solar cells.
    
        \subsubsection{Bosonic Optical Lattice}

            \begin{figure*}
                \centering
                \includegraphics[width=\linewidth]{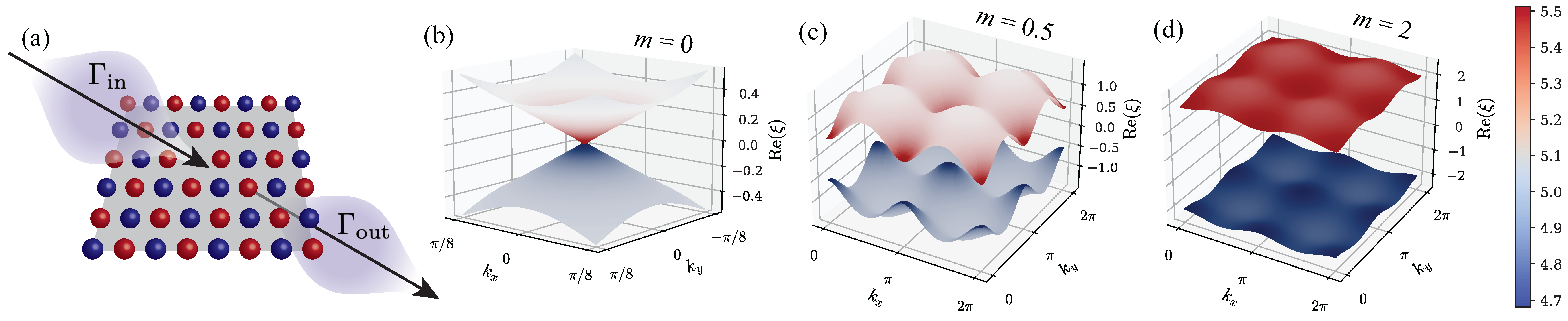}
                \caption{Occupation of a bipartite bosonic optical lattice with different trapping strengths on the top and the bottom and on the red (A) and blue (B) sublattices leads to a momentum-dependent filling. \textbf{(a)} Schematic illustration of an optical lattice subject to a cloud of atoms passing through the lattice providing a dissipative reservoir. \textbf{(b)} At $m=0$, the real part of the dispersion is sinusoidal and density fluctuations are most pronounced near the Dirac point. \textbf{(c)} As the gap opens density fluctuations spread through the Brillouin zone. \textbf{(d)} At large $m$, the density of the two bands become uniform over the zone. Plot parameters are $t=1$, $\Gamma_A^-=1.05$, $\Gamma_B^-=1.06$, $\Gamma_A^+=0.95$, $\Gamma_B^+=0.94$.}
                \label{fig:bosons}
            \end{figure*}

            \begin{figure}
                \centering
                \includegraphics[width=
                \linewidth]{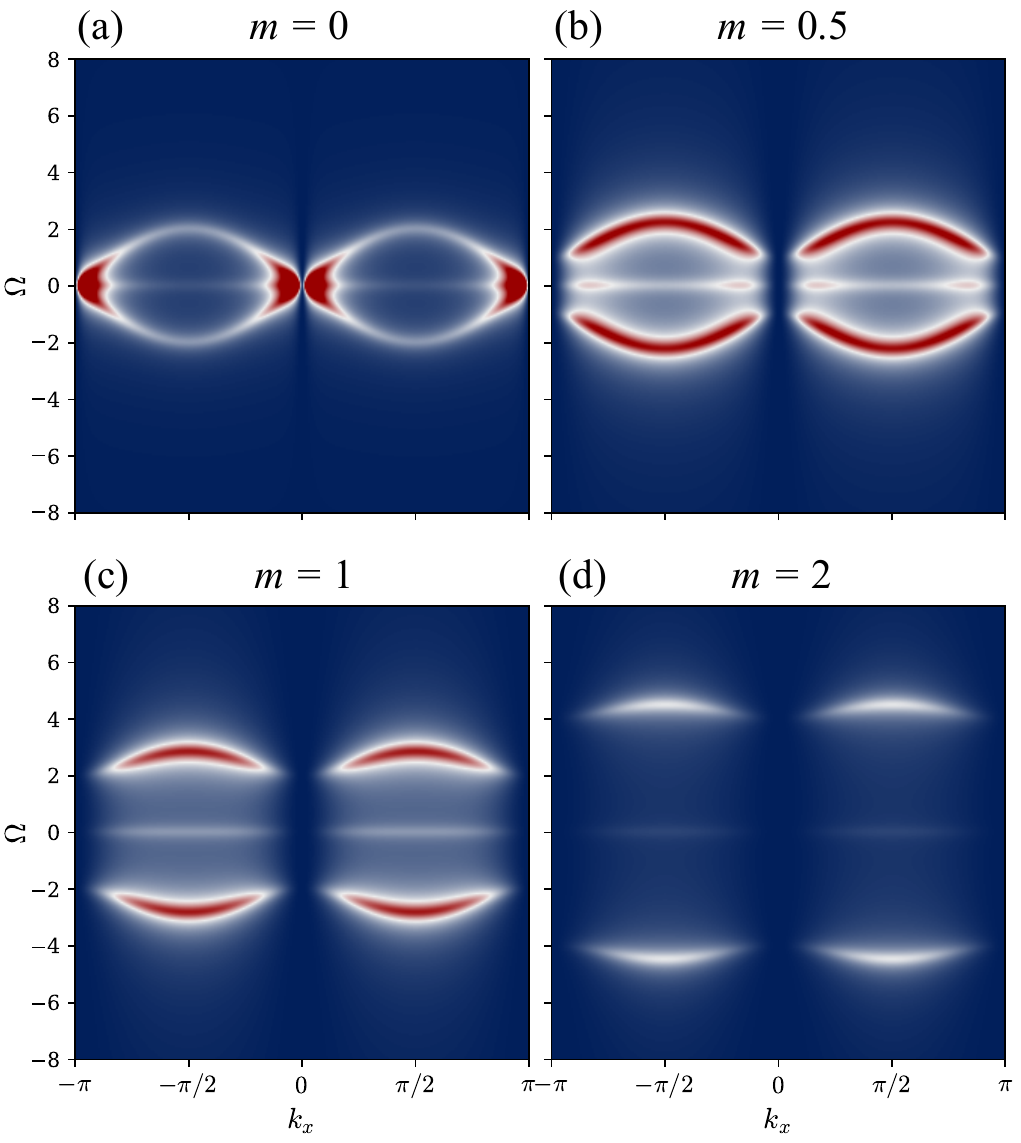}
                \caption{Magnitude of inter-orbital density-density correlation, $|\langle n_0 n_1\rangle(\Omega)|$ on a slice through $k$-space along $k_x$ with $k_y=0$. There is always a peak at $\Omega=0$ corresponding absorption followed by emission leading. Note that the structure is even about $\Omega=0$ corresponding to the presence of both stimulated absorption and stimulated emission processes. Additionally here we have chosen an inversion symmetric system with $\bm{k}$-independent jump operators so the correlation is also even about $\bm{k}=0$. \textbf{(a)} For $m=0$ the system is gapless and the correlation function peaks near the crossing but vanishes exactly at the crossing. \textbf{(b-c)} As the mass term increases the gap opens and the range over which the correlation function vanishes near the avoided crossing increases. \textbf{(d)} At large mass terms the correlation function diminishes as the mass term drives the bands to distinct limits of atomic orbitals.}
                \label{fig:bose-dynamic}
            \end{figure}

            Optical lattices are a tunable platform to realize exotic physics \cite{jaksch1998cold,bloch2008many}, and can be used to realize arbitrary lattices in two dimensions \cite{bloch2005ultracold,windpassinger2013engineering}. While optical lattices may approximate closed fermionic and bosonic systems, they are suspended in a near vacuum and are susceptible to loss and gain processes making dissipation fundamentally relevant for a full description of their physics \cite{pichler2010nonequilibrium,sharma2021driven}. Here we consider non-interacting bosons in an optical lattice subject to gain and loss processes.

            Here we consider a tight-binding model on a square lattice with nearest-neighbor hopping $t$ and a sublattice dependent mass term $m$
            \begin{align}
                H(k_x,k_y) = t\sin(k_x)\sigma_x + t\sin(k_y)\sigma_y + m \sigma_z,
            \end{align}
            where $\sigma_i$ are the Pauli matrices. We consider dissipation that acts uniformly on the sublattices with dissipators
            \begin{align}
                J^+_{A/B} &= \Gamma^+_{A/B} c^\dagger_{A/B}\\
                J^-_{A/B} &= \Gamma^-_{A/B} c_{A/B},
            \end{align}
            where $\Gamma$ are scalars describing the amplitude of the processes and $A$ and $B$ refer to the two sublattices.
            The anisotropy between gain and loss rates and between sublattices could correspond to the physical scenario of a stronger confining field on the top side and a weaker confining field on the bottom side and a bilayer with $A$ sites on the top layer and $B$ sites on the bottom layer. We illustrate such a process in Fig. \ref{fig:bosons}(a) where a cloud of atoms is incident from the upper left and leaves to the lower right.
    
            Using this model we can obtain the steady state density using the expressions from Section \ref{sec:density}. We find that at large $m$ the gap is large and the density of bosons is relatively uniform across the Brillouin zone. In contrast at small $m$, the density varies across the zone and varies rapidly near the Dirac point/band gap minima. We illustrate this in Fig. \ref{fig:bosons}(b-d) where the $z$-coordinate corresponds to the real part of the energy $\xi$ and the coloration corresponds to the density of bosons at that momentum in the optical lattice.

            We now study the finite-frequency correlation functions of the bosonic optical lattice. Using Eq. (\ref{eq:para_response}) with $O=n_i$ on orbital $i=0,1$ for the interorbital correlation function which we plot in Fig. \ref{fig:bose-dynamic}, where we have taken the magnitude to show the scale of the response, but not its phase. In the large mass limit, Fig. \ref{fig:bose-dynamic}(d), the bands are essentially the orbitals and the inter-orbital correlation is small. Meanwhile in the small mass limit, \ref{fig:bose-dynamic}(a), the bands are strong admixtures of the orbitals and the inter-orbital correlation is large.
            
    \section{Discussion}

        Quadratic Lindbladian systems present a generalization of non-interacting closed quantum systems to include dissipation, leading to dissipative steady states with new non-equilibrium features. Understanding their spectral properties and physical responses is essential to interpret realizations in quantum devices or engineered solid state systems. Here we systematically developed a Lindblad-Keldysh formalism to compute linear and non-linear dynamical response properties. In quadratic Lindbladian systems, we showed that the response probes the spectrum of an effective non-Hermitian Hamiltonian for second-quantized fermion or boson superoperators. The formalism  naturally applies to spin, electronic, and bosonic systems and their frequency dependent dynamic response properties, with generalizations to weakly interacting dissipative quantum systems an intriguing direction for future work.

        We first focus on the 1D XY spin model in a transverse field with boundary dissipation realized via jumps to two oppositely-polarized ferromagnetic reservoirs, which exhibits a non-equilibrium quantum phase transition between spin-density wave and paramagnetic phases \cite{prosen2008quantum,eisert2010noise}. In this setting, we studied the dynamical spin susceptibility at and away from criticality, which exhibit markedly distinct regimes of response as a function the boundary dissipation strengths. In contrast to the closed-system transverse field XY model, the dissipative steady state exhibits a series of gapless dispersive modes that originate from the underlying spin density wave pattern and disperse from $\pm 2q_\mathrm{SDW}~\mathrm{mod}~2\pi$. Additionally, spin-density wave amplitude excitations are present at $\pm q_\mathrm{SDW}$. The weak and ultra-strong dissipation limits exhibit strikingly similar spectra with an emergent inversion symmetry in the bulk and a correspondence of the dominant modes. This behavior originates from the fact that in both cases the effects of dissipation are confined to the boundary of the system and the bulk is relatively unaffected. In contrast, at commmensurate dissipation strength $\Gamma \sim J$, inversion symmetry is strongly broken and the asymmetry of dissipation at the two boundaries remains pronounced deep within the bulk. Finally, the response to local spin flips is most prominent at small fields, with excitations becoming more extended when the spin-density wavelength is large. With the dissipative Ising model recently realized in a quantum device \cite{mi2023stable}, measuring the unconventional excitations predicted by this work are a promising direction for fingerprinting out-of-equilibrium ordered phases.

        Turning to fermionic dissipative systems by example of bilayer graphene coupled to two reservoirs, we found that the diamagnetic optical conductivity can decrease with increasing occupation in stark contrast to closed systems; additionally for non-Hermitian current operators the diamagnetic response is dissipative rather than purely reactive. Moreover, second-order non-linear responses that are forbidden in the closed system by centrosymmetry emerge naturally as consequences of anisotropic coupling to a reservoir that lead to a non-centrosymmetric steady state. While we studied the AB stacked Bernal bilayer, the results are most reasonably obtained in large-twist angle twisted bilayer graphene where the energy of the quadratic band touching is drastically reduced \cite{mele2010commensuration,talkington2023electric}.

        Finally, in a bosonic optical lattice we found that sublattice-dependent gain and loss processes, for instance from sublattices living in different layers with different confining potentials, can lead to a steady state with a momentum-dependent occupation. When a sublattice dependent mass term is applied, the density in each band becomes uniform across the zone, in analogy to the equilibrium expectation of uniform occupation. Meanwhile in the inter-orbital dynamic response, the orbital-resolved response is small when the mass term is large and bands are well-approximated by atomic orbitals; in contrast the orbital-resolved response is large when the bands are mixtures of atomic orbitals.

        This work has a number of direct applications and possible generalizations. First, the computation of dynamical and spectroscopic response properties is essential to interpret dissipative steady states in quantum circuit realizations of quadratic Lindbladians on NISQ devices \cite{mi2023stable}. The developed formalism permits calculating response properties of arbitrary fermionic and bosonic systems that are described by a quadratic Lindbladian, making it immediately generalizable to weakly-interacting systems via perturbative expansions in Coulomb scattering. With rich phenomenologies established in closed interacting electron systems such as correlated metals \cite{maslov2016optical,kozii2017non} it will be interesting to study the ramifications of dissipation on their responses in an open system setting. Furthermore, charge or magnetic ordering transitions in dissipative interacting electron or boson systems are now simulable in cold atomic gases or quantum devices, necessitating dynamical probes of their properties. It will be interesting to generalize the presented formalism to mean field descriptions of dissipative ordered steady states, with applications to interacting photons and Rydberg atom arrays \cite{PhysRevA.84.031402,RevModPhys.85.299,weimer2021simulation}. Complementarily, driven-dissipative Floquet systems are known to exhibit interesting steady states and phase transitions with no thermal and closed-system counterpart \cite{PhysRevLett.116.070407,sieberer2023universality,PhysRevLett.122.110602}. Probing their electromagnetic response would serve as a useful diagnostic of engineering non-equilibrium response properties. Furthermore, with the role of quantum geometry and topology recently under much scrutiny in resonant responses of free electron systems \cite{morimoto2016topological,de2017quantized,ahn2022riemannian,onishi2023quantum} and low-frequency responses of strongly-interacting electron systems \cite{tai23}, it will be interesting to understand analogous signatures in dissipative responses in a non-Hermitian Lindbladian setting.  Here, an interesting question is how to devise experimentally accessible signatures of Lindbladian topology \cite{lieu2020tenfold}. Additionally, following this work, key areas of interest include signatures of non-Hermitian exceptional points in Lindbladian systems \cite{bergholtz2021exceptional}.

    \section*{Data Availability}

        The datasets generated during and/or analyzed during the current study are available from the corresponding author on reasonable request.

    \section*{Code Availability}

        The codes used during the current study are available from the corresponding author on reasonable request.

    \begin{acknowledgements}
        We thank Foster Thompson, Andrew Pocklington, and Pok Man Tam for discussions surrounding this work and comments on the manuscript.
        S.T. acknowledges support from the NSF under Grant No. DGE-1845298. M.C. acknowledges support from the NSF under Grant No. DMR-2132591.
    \end{acknowledgements}

    \section*{Author Contributions}

        S.T. oversaw the writing, implementation, and use of computer codes for the numeric simulation results present in this work. S.T. and M.C. both worked on the analytic derivations presented in this work. M.C. planned the study design. All authors were involved in the writing and revision of the manuscript. All authors have read and approved the manuscript. 

    \section*{Competing Interests}
    
        The authors declare no competing interests.

    \input{references.bbl}

	\newpage
	
    
	    \begin{widetext}
    
	    \supplement
    
	    \begin{center}
	    \textbf{\large Supplementary Material:\\ Linear and Non-Linear Response of Quadratic Lindbladians}

	    \vspace{12 pt}

	    Spenser Talkington and Martin Claassen
    
	    \vspace{12 pt}
    
	        \begin{minipage}{6 in}
	        {\small\phantom{ }\quad In this supplemental material, we: (1) derive a Kubo formula for Lindblad systems; (2) derive expressions for the equal-time (diamagnetic) and linear response (paramagnetic) responses, and express in terms of a spectral representation; (3) we derive an expression for the second order ``triangle" diagram that emerges in non-linear response, and express this diagram in terms of a spectral representation.}
	        \end{minipage}
	    \end{center}

	        \section{Kubo Formula}

	            In equilibrium, we have the Kubo formula [1]
	            \begin{align}
	                \langle O(t)\rangle = \langle O\rangle_0 + i \int_0^t d\bar{t}\ \langle[H'(\bar{t}),O(t)]\rangle_0,
	            \end{align}
	            where time evolution is given by the von Neumann equation $\partial_t O = -i[H',O]$.
	            Now for the Lindbladian system we consider (in the Schr\"odinger picture)
	            \begin{align}
	                \langle O\rangle(t) &= \frac{1}{2}(\text{Tr}[O\rho(t)] + \text{Tr}[\rho(t) O])\\
	                &= \frac{1}{2}(\text{Tr}[O\mathcal{T}e^{i\int_0^t d\bar{t}\ \mathcal{L}(\bar{t})}[\rho(0)]] + \text{Tr}[\mathcal{T}e^{i\int_0^t d\bar{t}\ \mathcal{L}(\bar{t})}[\rho(0)] O])
	            \end{align}
	            where we take the symmetric combination to result in a more symmetric final result, and $\mathcal{T}$ is the time-ordering operator. Now, we cannot directly expand the exponential since it is a superoperator rather than an operator, so we vectorize to obtain
	            \begin{align}
	                \langle O\rangle(t) &= \frac{1}{2}(\vec{\bm{1}}\cdot O_\ell\ \mathcal{T} e^{i\int_0^t d\bar{t} \hat{\mathcal{L}}(\bar{t})}\cdot\vec{\rho} + \vec{\bm{1}}\cdot O_r\ \mathcal{T} e^{i\int_0^t d\bar{t} \hat{\mathcal{L}}(\bar{t})}\cdot\vec{\rho})
	            \end{align}
	            where $O_\ell = \bm{1}\otimes O^\top$ and $O_r=O\otimes\bm{1}$. We can now transform to the interaction picture where we take $\hat{\mathcal{L}}=\hat{\mathcal{L}}_0+\sqrt{2}\hat{\mathcal{L}}' f(t)$ and $\hat{\hat{\mathcal{L}}}(\bar{t})=e^{i\hat{\mathcal{L}}_0\bar{t}}\hat{\mathcal{L}}'e^{-i\hat{\mathcal{L}}_0\bar{t}}$, so that
	            \begin{align}
	                \langle O\rangle(t) &= \frac{1}{2}\ \vec{\bm{1}}\cdot (O_\ell+O_r)\ e^{-i\hat{\mathcal{L}}_0 t} \mathcal{T} e^{\sqrt{2}i\int_0^t d\bar{t}\ \hat{\hat{\mathcal{L}}}' f(\bar{t})}\cdot\vec{\rho}
	            \end{align}
	            where we are now in a position to expand the matrix exponential in powers of $f$ which we assume to be small. Doing so to linear order we obtain
	            \begin{align}
	                \langle O\rangle(t) &= \frac{1}{2}\left( \vec{\bm{1}}\cdot (O_\ell+O_r)\ e^{-i\hat{\mathcal{L}}_0 t} (1+\sqrt{2}i\int_0^td\bar{t}\ e^{i\hat{\mathcal{L}}_0\bar{t}}\hat{\mathcal{L}}'e^{-i\hat{\mathcal{L}}_0\bar{t}} f(\bar{t}))\cdot\vec{\rho}\right)
	            \end{align}
	            Now, the zeroth order terms are traces over the steady state density matrix which vanish in the equilibrium setting, but are not guaranteed to vanish here since there may be a steady-state current. Separating terms we have
	            \begin{align}\label{eq:kubo-ness}
	                \langle O\rangle(t) &= \frac{1}{2}\left(\text{Tr}[(O_\ell+O_r) \rho_{ss}] + \vec{\bm{1}}\cdot (O_\ell+O_r)\ \sqrt{2}i\int_0^td\bar{t}\ e^{-i\hat{\mathcal{L}}_0(t-\bar{t})}\hat{\mathcal{L}}'e^{-i\hat{\mathcal{L}}_0\bar{t}} f(\bar{t})\cdot\vec{\rho}_{ss}\right)
	            \end{align}
	            We can now package this in a manner similar to the closed-system case by introducing $O_{c}=(O_\ell+O_r)/\sqrt{2}$ and $O_q=(O_\ell-O_r)/\sqrt{2}$ which for bosons are the ``classical" and ``quantum" fields and for fermions are just a rotation of frames. We then have
	            \begin{align}
	            \langle O\rangle(t) = \text{Tr}[\frac{1}{\sqrt{2}}\rho_{ss}O_c] + i \int_0^t d\bar{t} \langle O_c \hat{\mathcal{L}}'f(\bar{t}) \rangle_{ss}
	            \end{align}
	            where $\langle A \rangle_{ss} = \vec{\bm{1}}\cdot A \cdot \vec{\rho}_{ss}$.
	            See the next section of the Supplemental Material, for derivations of generic current operators $\hat{\mathcal{L}}'$ in terms of microscopic system parameters and the applied perturbation. Typically we will take $\hat{\mathcal{L}}' = O_q$, which corresponds to Hamiltonian perturbations. For dissipative perturbations, other forms of $\hat{\mathcal{L}}'$ are possible.

	            For clarity in the main text of this work we will consider Hamiltonian perturbations that act only on the system and \textit{not} on the couplings or the reservoir. We justify this limit by assuming that the relaxation time of the reservoir is much faster than the system (Markovian approximation) so that perturbations to the reservoir are irrelevant, and that the perturbations system remains relatively close to its steady state (linear response) so the couplings change immaterially with weak driving.

	            In this case, $O$ is a current operator given in the system, such as
	            \begin{align}
	                O^{\nu} = \mathsf{J}^\nu,
	            \end{align}
	            and $\mathcal{L}'[O(t)]=[J^\mu(t)A_\mu(t),O^\nu(0)]$
	            for the paramagnetic optical conductivity, where $\mathsf{J}^\mu=\partial H(t)/\partial k_\mu$ is entirely in the system.

	\section{Linear Response}
	
		In this section we consider linear response originating from equal-time (diamagnetic) and linear response (paramagnetic) bubbles.

	    \subsection{Equal-Time Response (Diamagnetic Response)}

	        Let us consider
	        \begin{align}
	            \mathsf{O} = \sum_{\bm{k},\alpha,\beta} O_{\bm{k},\alpha\beta} c_{\bm{k},\alpha}^\dagger c_{\bm{k},\beta}
	        \end{align}
	        where for the time-being we suppress the $\bm{k}$ index since $\bm{k}$ is associated to $\alpha$ and $\beta$. Now we have the correlation function
	        \begin{align}
	            \Pi_\text{dia}(t) &= -i\langle \mathsf{O}(t)\rangle\\
	            &= -i \sum_{\alpha,\beta} O_{\alpha\beta}\langle c_\alpha^\dagger(t) c_\beta(t) \rangle
	        \end{align}
	        where the response is instantaneous.
        
	        Now we recall from the density section, that
	        \begin{align}
	            \langle c_\alpha^\dagger(t)c_\beta(t)\rangle = \frac{e^{i\phi}}{2}\left(\delta_{\beta\alpha} - iG_{\beta\alpha}^K(t,t)\right)
	        \end{align}
	        where $\phi=\pi$ for fermions and $2\pi$ for bosons.
	        From this we have
	        \begin{align}
	            \Pi_\text{dia}(t) &= -i e^{i\phi}\sum_{\alpha,\beta} O_{\alpha\beta}\frac{1}{2}\left(\delta_{\beta\alpha} - iG_{\beta\alpha}^K(t,t)\right)\\
	            &= -\frac{i}{2} e^{i\phi}\bigg(\text{Tr}[O] - i\text{Tr}[OG_{\beta\alpha}^K(t,t)]\bigg)
	        \end{align}

	        Transforming to frequency space we have
	        \begin{align}
	            \Pi_\text{dia}(t) = -  \frac{i}{2}e^{i\phi} \bigg( \text{Tr}[O] -i \int_{-\infty}^{\infty} \frac{d\omega}{2\pi} e^{i\omega(t-t)}\ \text{Tr}[O G^K(\omega)] \bigg)
	        \end{align}
	        We can now expand using the spectral representation for $G^K$
	        \begin{align}
	            \Pi_\text{dia} = -  \frac{i}{2}e^{i\phi} \bigg( \text{Tr}[O] -i \int_{-\infty}^{\infty} \frac{d\omega}{2\pi}\ \sum_{n,n'} \frac{\langle\bar{u}_n|\sigma^i\Sigma^K\sigma^i|\bar{u}_{n'}\rangle}{(\omega - \xi_n)(\omega-\xi_{n'}^*)} \langle u_{n'}|O|u_n\rangle\bigg)
	        \end{align}
	        Completing the contour integral
	        \begin{align}
	            \Pi_\text{dia}
	            &= -  \frac{i}{2}e^{i\phi} \bigg( \text{Tr}[O] + \sum_{n,n'} \frac{\langle\bar{u}_n|\sigma^i\Sigma^K\sigma^i|\bar{u}_{n'}\rangle}{\xi_{n'}^* - \xi_n} \langle u_{n'}|O|u_n\rangle\bigg)
	        \end{align}
	        which is expressed in terms of the modes of $\sigma^i \Xi$.
    
	    \subsection{Linear Response (Paramagnetic Response)}

	        Let us consider
	        \begin{align}
	            \mathsf{O} = \sum_{\bm{k},\alpha,\beta} O_{\bm{k},\alpha\beta} c_{\bm{k},\alpha}^\dagger c_{\bm{k},\beta}
	        \end{align}
	        where for the time-being we suppress the $\bm{k}$ index since $\bm{k}$ is associated to $\alpha$ and $\beta$ while $\bm{k}'$ is associated to $\alpha'$ and $\beta'$. Now we have the correlation function
	        \begin{align}
	            \Pi_\text{para}(t) &= -i\langle [\mathsf{O}(t),\mathsf{O}(0)]\rangle \theta(t)\\
	            &= -i \sum_{\alpha,\beta, \alpha',\beta'}
	            O_{\alpha\beta} O_{\alpha'\beta'}\bigg(\langle c_{\alpha}^\dagger(t)c_{\beta}(t)c_{\alpha'}^\dagger(0)c_{\beta'}(0)\rangle - \langle c_{\alpha'}^\dagger(0)c_{\beta'}(0) c_{\alpha}^\dagger(t)c_{\beta}(t)\rangle\bigg)\theta(t)\\
	            &= -i \sum_{\alpha,\beta, \alpha',\beta'}
	            O_{\alpha\beta} O_{\alpha'\beta'}\bigg(\langle c_{\alpha}^\dagger(t)c_{\beta'}(0)\rangle\langle c_{\beta}(t)c_{\alpha'}^\dagger(0)\rangle - \langle c_{\alpha'}^\dagger(0)c_{\beta}(t)\rangle\langle c_{\beta'}(0) c_{\alpha}^\dagger(t)\rangle\bigg)\theta(t)
	        \end{align}
	        where we have completed the Wick contraction

	            \begin{align}\label{eq:wick-4}
	                \langle \alpha_1 \alpha_2 \alpha_3 \alpha_4\rangle
	                = \langle \alpha_1 \alpha_2\rangle\langle\alpha_3 \alpha_4\rangle + e^{i\phi}
	                \langle\alpha_1 \alpha_3\rangle\langle\alpha_2 \alpha_4\rangle + \langle \alpha_1 \alpha_4\rangle\langle \alpha_2 \alpha_3\rangle.
	            \end{align}
        
	        Expressing in terms of Green's functions 
	        \begin{align}
	            G_{\alpha\beta}^<(0,t) = - ie^{i\phi}\langle c^\dagger_{\beta}(0)c_{\alpha}(t)\rangle\\
	            G_{\alpha\beta}^>(0,t) = - i\langle c_{\alpha}(t)c^\dagger_{\beta}(0)\rangle
	        \end{align}
	        we have
	        \begin{align}
	            \Pi_\text{para}(t)
	            &= i e^{i\phi} \sum_{\alpha,\beta, \alpha',\beta'}
	            O_{\alpha\beta} O_{\alpha'\beta'}\bigg(G^<_{\beta'\alpha}(t,0) G^>_{\beta\alpha'}(0,t) - G^>_{\beta'\alpha}(t,0)G^<_{\beta\alpha'}(0,t) \bigg)\theta(t)
	        \end{align}
        
	        Reexpressing $G^<$ and $G^>$ in terms of $G^R$, $G^A$ and $G^K$ as in Eq. (\ref{eq:Grak}) of the main text we find
	        \begin{align}
	            \Pi_\text{para}(t)
	            &= \frac{i}{2}e^{i\phi} \bigg(\text{Tr}[O G^R(0,t)O' G^K(t,0)]+\text{Tr}[O G^K(0,t)O' G^A(t,0)]\bigg)
	        \end{align}
    
	        Now, we can take the Fourier transform of this to find
	        \begin{align}
	            \Pi_\text{para}(\Omega) = \frac{i}{2} e^{i\phi}\int_{-\infty}^\infty \frac{d\omega}{2\pi}\ \bigg(\text{Tr}[O G^R(\omega)O' G^K(\omega+\Omega)]+\text{Tr}[O G^K(\omega-\Omega)O' G^A(\omega)]\bigg)
	        \end{align}
	        We can evaluate the integral using the residue theorem at each pole $\omega_q$
	        \begin{align}
	            \Pi_\text{para}(\Omega) = -\frac{e^{i\phi}}{2}\bigg(\sum_q \text{Res}[\text{Tr}[O G^R(\omega_q)O' G^K(\omega_q+\Omega)],\omega_q] + \sum_q \text{Res}[\text{Tr}[O G^K(\omega_q-\Omega)O' G^A(\omega_q)],\omega_q]\bigg)
	        \end{align}

	        Now we have $G^K= G^R \Sigma^K G^A$, so
	        \begin{align}
	            \Pi_\text{para}(\Omega) = -e^{i\phi} &\sum_q \text{Res}[\text{Tr}[O G^R(\omega_q)O' G^R(\omega_q+\Omega)\Sigma^K G^A(\omega_q+\Omega)],\omega_q]\\ -e^{i\phi}&\sum_q \text{Res}[\text{Tr}[O G^R(\omega_q-\Omega)\Sigma^K G^A(\omega_q-\Omega)O' G^A(\omega_q)],\omega_q]
	        \end{align}
	        and we have
	        \begin{align}\label{eq:gf}
	            G^R(\omega_q) = \sum_n\frac{|u_n\rangle\langle\bar{u}_n|}{\omega_q-\xi_n} \sigma^i\\
	            G^A(\omega_q) = \sum_n \sigma^i\frac{|\bar{u}_n\rangle\langle u_n|}{\omega_q-\xi_n^*}
	        \end{align}
	        where the bars indicate \textit{left} eigenstates.
	        Inserting these expressions we then have
	        \begin{align}
	            \Pi_\text{para}(\Omega) = -e^{i\phi}&\sum_{q,n,n',n''} \text{Res}[ \frac{\langle u_{n''}|O|u_n\rangle\langle\bar{u}_n|\sigma^iO'|u_{n'}\rangle\langle\bar{u}_{n'}|\sigma^i\Sigma^K\sigma^i|\bar{u}_{n''}\rangle}{(\omega_q-\xi_n)(\omega_q+\Omega-\xi_{n'})(\omega_q+\Omega-\xi_{n''}^*)},\omega_n]\nonumber\\
	            -e^{i\phi}&\sum_{q,n,n',n''} \text{Res}[ \frac{\langle u_{n''}|O'\sigma^i|\bar{u}_{n}\rangle\langle u_{n}|O|u_{n'}\rangle\langle\bar{u}_{n'}|\sigma^i\Sigma^K\sigma^i|\bar{u}_{n''}\rangle}{(\omega_q-\xi_{n}^*)(\omega_q-\Omega-\xi_{n'})(\omega_q-\Omega-\xi_{n''}^*)},\omega_q]
	        \end{align}
	        closing upwards/downwards and taking the residue,
	        \begin{align}
	            \Pi_\text{para}(\Omega) = -e^{i\phi} &\sum_{n,n',n''} \frac{\langle u_{n''}|O|u_n\rangle\langle\bar{u}_n|\sigma^iO'|u_{n'}\rangle\langle\bar{u}_{n'}|\sigma^i\Sigma^K\sigma^i|\bar{u}_{n''}\rangle}{(\xi_{n''}^*-\xi_n-\Omega)(\xi_{n''}^*-\xi_{n'})} - \frac{\langle u_{n''}|O'\sigma^i|\bar{u}_{n}\rangle\langle u_{n}|O|u_{n'}\rangle\langle\bar{u}_{n'}|\sigma^i\Sigma^K\sigma^i|\bar{u}_{n''}\rangle}{(\xi_{n'}-\xi_{n}^*+\Omega)(\xi_{n'}-\xi_{n''}^*)}
	        \end{align}
	        and so the correlation function is 
	        \begin{align}
	            \Pi_\text{para}(\Omega) = -e^{i\phi}\sum_{n,n'} \frac{\langle\bar{u}_{n}|\sigma^i\Sigma^K\sigma^i|\bar{u}_{n'}\rangle}{\xi_{n'}^*-\xi_{n}}
	            \bigg(\sum_{n''}\frac{\langle u_{n'}|O|u_{n''}\rangle\langle\bar{u}_{n''}|\sigma^iO'|u_{n}\rangle}{(\xi_{n'}^*-\xi_{n''})-\Omega} + \frac{\langle u_{n'}|O'\sigma^i|\bar{u}_{n''}\rangle\langle u_{n''}|O|u_{n}\rangle}{(\xi_{n}-\xi_{n''}^*)+\Omega}\bigg)
	        \end{align}
	        where we reindexed for consistency with the diamagnetic part.

	\section{Non-Linear Response}

	        For the second order response we have the new ``triangle diagram" term
	        \begin{align}
	            \langle [\O(0),[\O'(t),\O''(t+t')]]\rangle\theta(t)\theta(t')
	            = &[\langle \O(0)\O'(t)\O''(t+t')\rangle - \langle \O(0)\O''(t+t')\O'(t)\rangle\nonumber\\ - &\langle \O'(t)\O''(t+t')\O(0)\rangle + \langle \O''(t+t')\O'(t)\O(0)\rangle]\theta(t)\theta(t')
	        \end{align}
	        where for perturbations to the system only we have
	        \begin{align}
	            \O = \sum_{\bm{k},\alpha,\beta} O_{\bm{k},\alpha\beta} c^\dagger_{\bm{k},\alpha}c_{\bm{k},\beta}
	        \end{align}
	        so we will be interested in terms like
	        \begin{align}
	            \sum_{\alpha,\beta,\alpha',\beta',\alpha'',\beta''} O_{\alpha\beta} O'_{\alpha'\beta'} O''_{\alpha''\beta''} \langle c^\dagger_{\alpha}(t)c_{\beta}(t)c^\dagger_{\alpha'}(t')c_{\beta'}(t')c^\dagger_{\alpha''}(t'')c_{\beta''}(t'')\rangle
	        \end{align}
	        where we suppress the $\bm{k}$ indices. This can then be expanded in terms of two point functions using Wick's theorem [2].
	        For six fermions/bosons we have
	        \begin{align}
	        	\langle \alpha_1 \alpha_2 \alpha_3 \alpha_4 \alpha_5 \alpha_6\rangle
	        	= &\langle \alpha_1 \alpha_2\rangle \langle\alpha_3 \alpha_4 \alpha_5 \alpha_6\rangle\nonumber\\
	        	+ &e^{i\phi}\langle\alpha_1 \alpha_3\rangle \langle\alpha_2 \alpha_4 \alpha_5 \alpha_6\rangle\nonumber\\
	        	+ &\langle\alpha_1 \alpha_4\rangle \langle\alpha_2 \alpha_3 \alpha_5 \alpha_6\rangle\nonumber\\
	        	+ &e^{i\phi}\langle\alpha_1 \alpha_5\rangle \langle\alpha_2 \alpha_3 \alpha_4 \alpha_6\rangle\nonumber\\
	        	+ &\langle \alpha_1 \alpha_6\rangle \langle\alpha_2 \alpha_3 \alpha_4 \alpha_5\rangle,
	        \end{align}
	        where $\phi=\pi$ for fermions and $2\pi$ for bosons.
	        As above, we can reexpress these two-point functions in terms of $G^R$, $G^A$ and $G^K$.
	        For $T>t$ we have $G^R(T,t)=0$ and $G^A(t,T)=0$ which can be used to simplify the result.
	        Completing the Wick contraction and expression in terms of Green's functions and writing as a trace, we have
	        \begin{align}
	            \langle [\O(0),[\O'(t),\O''(t+t')]]\rangle\theta(t)\theta(t') =
	            -\frac{i}{2}e^{i\phi}
	            \bigg(& \text{Tr}[O G^K(0,t+t') O'' G^A(t+t',t) O' G^A(t,0)]  \nonumber\\
	            +& \text{Tr}[O G^R(0,t+t') O'' G^A(t+t',t) O' G^K(t,0)] \nonumber\\
	            +& \text{Tr}[O G^K(0,t) O' G^R(t,t+t') O'' G^A(t+t',0)]\nonumber\\
	            +& \text{Tr}[O G^R(0,t) O' G^R(t,t+t') O'' G^K(t+t',0)] \bigg)
	        \end{align}

	        \phantom{ }

	        Transforming to frequency space, we have
	        \begin{align}
	            \langle [\O(\Omega),[\O'(\Omega'),\O''(\Omega+\Omega')]]\rangle = -\frac{i}{2}e^{i\phi} \int \frac{d\omega}{2\pi}~ \bigg( &\Tr[ O G^K(\omega + \Omega + \Omega') O'' G^A(\omega) O' G^A(\omega + \Omega') ] \notag\\
	        	+& \Tr[ O G^R(\omega + \Omega + \Omega') O'' G^A(\omega) O' G^K(\omega + \Omega') ] \notag\\
	        	+& \Tr[ O G^K(\omega - \Omega') O' G^R(\omega) O'' G^A(\omega - \Omega - \Omega') ]  \notag\\
	        	+& \Tr[ O G^R(\omega - \Omega') O' G^R(\omega) O'' G^K(\omega - \Omega - \Omega') ]  \bigg)
	        \end{align}
	        We now substitute the Keldysh Green's function
	        \begin{align}
	        	G^K(\omega) = G^R(\omega) \Sigma^K G^A(\omega)
	        \end{align}
	        to obtain
	        \begin{align}
	        	\langle [\O(\Omega),[\O'(\Omega'),\O''(\Omega+\Omega')]]\rangle = -ie^{i\phi} \int \frac{d\omega}{2\pi} \{ &\Tr[ O G^R(\omega + \Omega + \Omega') \Sigma^K G^A(\omega + \Omega + \Omega')  O'' G^A(\omega) O' G^A(\omega + \Omega') ] \notag\\
	        	+& \Tr[ O G^R(\omega + \Omega + \Omega') O'' G^A(\omega) O' G^R(\omega + \Omega') \Sigma^K G^A(\omega + \Omega') ] \notag\\
	        	+& \Tr[ O G^R(\omega - \Omega') \Sigma^K G^A(\omega - \Omega') O' G^R(\omega) O'' G^A(\omega - \Omega - \Omega') ]  \notag\\
	        	+& \Tr[ O G^R(\omega - \Omega') O' G^R(\omega) O'' G^R(\omega - \Omega - \Omega') \Sigma^K G^A(\omega - \Omega - \Omega') ]  \}
	        \end{align}
	        expanding using the spectral representation, Eq. (\ref{eq:gf}), we obtain
	        \begin{align}
	        	\langle [\O(\Omega),[\O'(\Omega'),\O''(\Omega+\Omega')]]\rangle = -ie^{i\phi} \int \frac{d\omega}{2\pi} \sum_{n_1\cdots n_4} &\bigg( \frac{ \braOPket{u_{n_4}}{O}{u_{n_1}}  \braOPket{\bar{u}_{n_1}}{\sigma^i\Sigma^K\sigma^i}{\bar{u}_{n_2}} \braOPket{u_{n_2}}{O''\sigma^i}{\bar{u}_{n_3}} \braOPket{u_{n_3}}{O'\sigma^i}{\bar{u}_{n_4}} }{ (\omega+\Omega+\Omega' - \E_{n_1}) (\omega+\Omega+\Omega' - \E^*_{n_2}) (\omega - \E^*_{n_3})(\omega+\Omega' - \E^*_{n_4})} \notag\\
	        \
	        	&+ \frac{ \braOPket{u_{n_4}}{O}{u_{n_1}}  \braOPket{\bar{u}_{n_1}}{\sigma^i O'' \sigma^i}{\bar{u}_{n_2}} \braOPket{u_{n_2}}{O'}{u_{n_3}} \braOPket{\bar{u}_{n_3}}{\sigma^i\Sigma^K\sigma^i}{\bar{u}_{n_4}} }{ (\omega+\Omega+\Omega' - \E_{n_1}) (\omega - \E^*_{n_2}) (\omega+\Omega' - \E_{n_3})(\omega+\Omega' - \E^*_{n_4})} \notag\\
	        \
	        	&+ \frac{ \braOPket{u_{n_4}}{O}{u_{n_1}}  \braOPket{\bar{u}_{n_1}}{\sigma^i\Sigma^K\sigma^i}{\bar{u}_{n_2}} \braOPket{u_{n_2}}{O'}{u_{n_3}} \braOPket{\bar{u}_{n_3}}{\sigma^i O'' \sigma^i}{\bar{u}_{n_4}} }{ (\omega-\Omega' - \E_{n_1}) (\omega-\Omega' - \E^*_{n_2}) (\omega - \E_{n_3})(\omega-\Omega-\Omega' - \E^*_{n_4})} \notag\\
	        \
	        	&+ \frac{ \braOPket{u_{n_4}}{O}{u_{n_1}}  \braOPket{\bar{u}_{n_1}}{\sigma^i O'}{u_{n_2}} \braOPket{\bar{u}_{n_2}}{\sigma^i O''}{u_{n_3}} \braOPket{\bar{u}_{n_3}}{\sigma^i\Sigma^K\sigma^i}{\bar{u}_{n_4}} }{ (\omega-\Omega' - \E_{n_1}) (\omega - \E_{n_2}) (\omega - \Omega - \Omega' - \E_{n_3})(\omega - \Omega - \Omega' - \E^*_{n_4})} \bigg)
	        \end{align}
	        We can now evaluate the frequency integral, where we note that $\textrm{Im}(\E_m) < 0$. We find
	        \begin{align}
	        	\langle [\O(\Omega),[\O'(\Omega'),\O''(\Omega+\Omega')]]\rangle =  e^{i\phi}\sum_{n_1\cdots n_4} &\left\{ \frac{ \braOPket{u_{n_4}}{O}{u_{n_1}}  \braOPket{\bar{u}_{n_1}}{\sigma^i\Sigma^K\sigma^i}{\bar{u}_{n_2}} \braOPket{u_{n_2}}{O'' \sigma^i}{\bar{u}_{n_3}} \braOPket{u_{n_3}}{O' \sigma^i}{\bar{u}_{n_4}} }{ (\E_{n_1} - \E^*_{n_2}) (\Omega + \Omega' + \E^*_{n_3} - \E_{n_1}) (\Omega + \E^*_{n_4} - \E_{n_1})} \right. \notag\\
	        \
	        	&+ \braOPket{u_{n_4}}{O}{u_{n_1}}  \braOPket{\bar{u}_{n_1}}{\sigma^i O'' \sigma^i}{\bar{u}_{n_2}} \braOPket{u_{n_2}}{O'}{u_{n_3}} \braOPket{\bar{u}_{n_3}}{\sigma^i\Sigma^K\sigma^i}{\bar{u}_{n_4}} \times \notag\\
	        		&~~~~\times \bigg( \frac{1}{(\xi_{n_4}^*-\xi_{n_3})(\Omega'+\xi_{n_2}^*-\xi_{n_3})(\Omega+\xi_{n_3}-\xi_{n_1})}\notag\\ &\qquad -  \frac{1}{(\Omega+\Omega'+\xi_{n_2}^*-\xi_{n_1})(\Omega+\xi_{n_3}-\xi_{n_1})(\Omega+\xi_{n_4}^*-\xi_{n_1})} \bigg) \notag\\
	        \
	        	&+ \braOPket{u_{n_4}}{O}{u_{n_1}}  \braOPket{\bar{u}_{n_1}}{\sigma^i\Sigma^K\sigma^i}{\bar{u}_{n_2}} \braOPket{u_{n_2}}{O'}{u_{n_3}} \braOPket{\bar{u}_{n_3}}{\sigma^i O'' \sigma^i}{\bar{u}_{n_4}} \notag\\
	        	&~~~~\times \bigg( \frac{1}{(\xi_{n_2}^*-\xi_{n_1})(\Omega'+\xi_{n_1}-\xi_{n_3})(\Omega+\xi_{n_4}^*-\xi_{n_1})}\notag\\ &\qquad - \frac{1}{(\Omega'+\xi_{n_1}-\xi_{n_3})(\Omega'+\xi_{n_2}^*-\xi_{n_3})(\Omega+\Omega'+\xi_{n_4}^*-\xi_{n_3})} \bigg) \notag\\
	        \
	        	&- \left.\frac{ \braOPket{u_{n_4}}{O}{u_{n_1}}  \braOPket{\bar{u}_{n_1}}{\sigma^i O'}{u_{n_2}} \braOPket{\bar{u}_{n_2}}{\sigma^i O''}{u_{n_3}} \braOPket{\bar{u}_{n_3}}{\sigma^i\Sigma^K\sigma^i}{\bar{u}_{n_4}} }{ (\E_{n_3} - \E^*_{n_4})(\Omega + \E^*_{n_4} - \E_{n_1})(\Omega + \Omega' + \E^*_{n_4} - \E_{n_2})} \right\}
	        \end{align}
	        which is expressed in terms of the modes of $\sigma^i \Xi$.

	        \begin{center}
	        \textbf{SUPPLEMENTARY REFERENCES}
	    \end{center}

	    [1] R. Kubo. J. Phys. Soc. Japan \textbf{12}, 570 (1957).

	    [2] A. Altland and B. D. Simons, \textit{Condensed Matter Field Theory} (Cambridge University Press, 2010).

	    \newpage
        
	    \end{widetext}

\end{document}

%% file: references.bbl
%